\documentclass[11pt,a4paper]{article}
\pdfoutput=1

\usepackage{jheppub}
\usepackage{latexsym}
\usepackage{revsymb}
\usepackage{multirow}
\usepackage{color}
\usepackage[usenames,dvipsnames,svgnames,table]{xcolor}

\usepackage{graphicx}% Include  files
\usepackage{epsfig}  % Include figure files
\usepackage{epsf}    % Include figure files
\usepackage{dcolumn}% Align table columns on decimal point
\usepackage{bm}% bold math
\usepackage{textcomp}% Align table columns on decimal point
\usepackage{float}
\usepackage{hypcap}
\usepackage[]{hyperref}
\usepackage{adjustbox}
\usepackage[utf8]{inputenc}
\usepackage[T1]{fontenc} % if needed
\usepackage{url}
\usepackage{amsmath}
\usepackage{mathtools}
\usepackage{mathrsfs}
\usepackage{subfigure}
\usepackage{alphalph}
\usepackage{morefloats}
\usepackage{textcomp}
\usepackage{comment}

\colorlet{shadecolor}{gray!15}
\definecolor{greenLinks}{rgb}{0, 0.6, 0} 
\definecolor{blueLinks}{rgb}{0, 0, 0.6}
\definecolor{redLinks}{rgb}{0.6, 0, 0}
\definecolor{tempText}{rgb}{0.55, 0.10,0.67}
\definecolor{eprintLinks}{rgb}{0.4, 0.4, 0.4}
\definecolor{journalLinks}{rgb}{0.6, 0, 0}

\hypersetup{colorlinks=true,linkcolor=blueLinks,citecolor=blueLinks,urlcolor=blueLinks}

\newcommand\barparena[1]{\overset{\scriptscriptstyle(-)}{#1}}

\def\21{$\mathrm{SU(2)_L \otimes U(1)_Y}$ }

\definecolor{gray}{rgb}{0.5,0.5,0.5}

\title{Testing the heavy decaying sterile neutrino hypothesis at the DUNE near detector}

\author[a,b]{Sabya Sachi Chatterjee,}
\affiliation[a]{Institut f\"{u}r Astroteilchenphysik, Karlsruher Institut f\"{u}r Technologie (KIT), Hermann-von-Helmholtz-Platz 1, 76344 Eggenstein-Leopoldshafen, Germany}
\emailAdd{sabya.chatterjee@kit.edu}

\author[b]{St{\'e}phane Lavignac,}
\affiliation[b]{Institut de Physique Th{\'e}orique, Universit{\'e} Paris Saclay, CNRS, CEA, F-91191 Gif-sur-Yvette, France}
\emailAdd{stephane.lavignac@ipht.fr}

\author[c]{O. G. Miranda}
\affiliation[c]{Departamento de F\'{\i}sica, Centro de Investigaci\'on
  y de Estudios Avanzados del IPN,\\ Apartado Postal 14-740 07000,
  Ciudad de Mexico, Mexico}
\emailAdd{omar.miranda@cinvestav.mx}

\abstract{
One of the most convincing explanations of the LSND and MiniBooNE anomalies
relies on a heavy, mostly sterile neutrino with
a small muon neutrino component, which decays to an electron neutrino and an invisible light
scalar field. We investigate the possibility to test this hypothesis at the near detector complex
of the upcoming DUNE experiment. We find that the DUNE liquid argon near detector (ND-LAr)
can probe a larger region of the parameter space than the Fermilab SBN program,
and may help to confirm or reject a possible hint of $\nu_e$ appearance in future
MicroBooNE, SBND or ICARUS data.
We also argue that it may be possible to distinguish between Dirac and Majorana neutrinos
if this scenario is realized in Nature.
}
\begin{document}
\maketitle

%%%%%%%%%%%
\section{Introduction}
\label{sec:intro}
%%%%%%%%%%%

While most of the large amount of available data from solar, atmospheric, reactor and
accelerator neutrino experiments is very successfully interpreted in terms of 3-flavour
oscillations~\cite{Esteban:2024eli,deSalas:2020pgw,Capozzi:2021fjo}, a few anomalies
still persist. Among these, the oldest ones come from the LSND~\cite{LSND:2001aii} and
MiniBooNE~\cite{MiniBooNE:2020pnu} short-baseline accelerator experiments.
LSND, which ran from 1993 to 1998, observed a $3.8 \sigma$ excess of $\bar \nu_e$ events
in a $\bar \nu_\mu$ beam
that cannot be accounted for by oscillation parameters consistent with the other experiments.
This excess was neither confirmed nor excluded by the (less sensitive) KARMEN experiment~\cite{KARMEN:2002zcm}.
The goal of the MiniBooNE experiment was to test the LSND anomaly with a different
baseline ($541\, \mbox{m}$ vs. $30\, \mbox{m}$ for LSND) but a similar L/E. 
MiniBooNE ran from 2002 to 2019 with muon neutrinos and antineutrinos and observed
an excess of $\nu_e$ ($\bar \nu_e$) events at low energy in both modes, with a statistical
significance of $4.8 \sigma$~\cite{MiniBooNE:2020pnu} (the overall statistical significance
of the LSND and MiniBooNE anomalies is $6.1 \sigma$).
The MiniBooNE excess can be interpreted in terms of
$\barparena{\nu}_{\!\! \mu}\rightarrow \barparena{\nu}_{\!\! e}$ oscillations
involving a fourth, mostly sterile neutrino in the eV range, with parameters
$\Delta m^2 \gtrsim 0.1\, \mbox{eV}^2$ and $\sin^2 2 \theta \approx (10^{-3} - 10^{-2})$,
in good agreement with the LSND allowed region.
However, this explanation would imply $\nu_\mu$/$\bar \nu_\mu$ and $\nu_e$/$\bar \nu_e$
disappearance governed by the same $\Delta m^2$, which is not observed by
experiments such as MINOS/MINOS+~\cite{MINOS:2017cae}, IceCube DeepCore~\cite{IceCube:2024dlz},
and the reactor neutrino experiments Bugey~\cite{Declais:1994su} and Daya Bay~\cite{DayaBay:2024nip}.
This tension between appearance and disappearance data
has been quantified to be at $4.7\sigma$ level in Ref.~\cite{Dentler:2018sju}.
In addition, a sterile neutrino with parameters consistent with LSND and MiniBooNE
is strongly disfavoured by cosmology~\cite{Planck:2018vyg}.

Thus, if the LSND and MiniBooNE anomalies are real (i.e., not attributable to misunderstood
backgrounds or experimental errors), it appears unlikely that they are due to oscillations driven
by light sterile neutrinos, and an alternative, non-oscillation explanation is needed.
It is therefore crucial to establish whether these experimental anomalies are real or not.
This is the goal of the
short-baseline accelerator neutrino program at Fermilab (SBN)~\cite{MicroBooNE:2015bmn},
which consists of three liquid argon detectors with very good event reconstruction capabilities:
SBND~\cite{AlvarezGarrote:2024szs} (with a baseline of 110 m),
MicroBooNE~\cite{MicroBooNE:2016pwy} (470 m) and ICARUS~\cite{ICARUS:2023gpo} (600 m).
So far only MicroBooNE has published results, testing various possible origins
of the MiniBooNE low-energy excess\footnote{The MiniBooNE detector was not able
to efficiently distinguish electrons from photons and collimated $e^+ e^-$ pairs.},
such as single electron~\cite{MicroBooNE:2024ryw},
single photon~\cite{MicroBooNE:2025ntu} or $e^+ e^-$ pairs~\cite{MicroBooNE:2025khi}.
In particular, MicroBooNE recently excluded
the MiniBooNE low-energy excess as $\nu_e$ at more than $99\%$ C.L.~\cite{MicroBooNE:2024ryw}.

Still, more data is needed to fully exclude the MiniBooNE anomaly, and it makes sense
to investigate the possibility of testing non-oscillation explanations
of the LSND and MiniBooNE excesses at SBN or at other experimental facilities~\cite{Brdar:2020tle}.
Among the possible new physics explanations (see e.g. Ref.~\cite{Acero:2022wqg} for a review),
one of the most convincing is the heavy decaying sterile neutrino
initially proposed for LSND by the authors of Ref.~\cite{Palomares-Ruiz:2005zbh},
and extended to MiniBooNE in Refs.~\cite{Dentler:2019dhz} and~\cite{deGouvea:2019qre}.
In this scenario, a keV/MeV-scale sterile neutrino mixing with $\nu_\mu$ and decaying to $\nu_e$
and a light scalar field
is able to mimic the LSND and MiniBooNE excesses even if its mixing angles with $\nu_\mu$ and $\nu_e$
are small, thus avoiding the tension with appearance experiments
(for other proposals involving a heavy sterile neutrino decaying to visible or invisible final states, see
Refs.~\cite{Gninenko:2009ks,Gninenko:2010pr,Masip:2012ke,Bai:2015ztj,Magill:2018jla,Bertuzzo:2018itn,Ballett:2018ynz,Ballett:2019pyw,Fischer:2019fbw,Moulai:2019gpi,Datta:2020auq,Dutta:2020scq,Abdallah:2020biq,Abdullahi:2020nyr,Abdallah:2020vgg,Chang:2021myh,Vergani:2021tgc,Hammad:2021mpl}).

The goal of this paper is to investigate the possibility to test the heavy decaying sterile neutrino (HDSN)
hypothesis at the DUNE liquid argon near detector (ND-LAr), focusing for definiteness on the scenario of Ref.~\cite{deGouvea:2019qre}.
Our main result is that ND-LAr can probe a larger portion of the HDSN parameter space
than the Fermilab SBN program,
which might prove crucial to check a possible hint of $\nu_e$ appearance
in future MicroBooNE, SBND or ICARUS data.
We also find that it may be possible, in case of a positive signal,
to distinguish between Dirac and Majorana neutrinos.

The paper is organized as follows. In Section~\ref{sec:NuDecay}, we introduce the heavy decaying
sterile neutrino scenario considered in this paper and provide the main formulae
used in our analysis. Section~\ref{sec:experiment} briefly presents the specifications of the DUNE
liquid argon near detector and describes the simulations performed in our paper.
In Section~\ref{sec:sensitivity}, we present our results for the expected sensitivity of ND-LAr
to the HDSN parameters, and comment about the possibility to distinguish between Dirac
and Majorana neutrinos if this scenario is realized in Nature.
Finally, we give our conclusions in Section~\ref{sec:conclusions}.

%%%%%%%%%%%%%%%%%%
\section{The heavy decaying sterile neutrino (HDSN) scenario}
\label{sec:NuDecay}
%%%%%%%%%%%%%%%%%%

In this work, we consider the heavy decaying sterile neutrino scenario of Ref.~\cite{deGouvea:2019qre},
hereafter referred to as the HDSN scenario.
It involves a fourth, mostly sterile neutrino mass eigenstate $\nu_4$
with a mass in the keV-MeV range and a small $\nu_\mu$ component. This heavy neutrino decays to an electron neutrino
and an invisible light scalar field $\phi$, thus mimicking the excesses observed by LSND and MiniBooNE.
More specifically, the sterile neutrino is assumed to mix only with the muon neutrino:
\begin{equation}
  \nu_{\alpha L}\, =\, \sum_{i=1}^4 U_{\alpha i} \nu_{i L}\, , \qquad U_{e4} = U_{\tau 4} = 0\, ,
\label{eq:sterile_active_mixing}
\end{equation}
where $\alpha = e, \mu, \tau, s$ and $U$ is the $4 \times 4$ lepton mixing matrix.
The low-energy effective Lagrangian responsible for the decay of $\nu_4$ is taken
to be~\cite{Palomares-Ruiz:2005zbh,deGouvea:2019qre}
\begin{equation}
%  {\cal L}\, =\, - g\, \bar \nu_4 \nu_{eL} \phi + \mbox{h.c.}\, .
  {\cal L}_{\rm decay}\, =\, - g\, \bar \nu_{4R} \nu_{eL} \phi + \mbox{h.c.}\, .
\label{eq:interaction}
\end{equation}
If neutrinos are Dirac fermions, this corresponds to a maximally parity violating
interaction involving only $\nu_{4R}$ and $\nu_{eL}$, while in the Majorana case $\nu_{4R}$
is the CP conjugate of $\nu_{4L}$ (i.e., $\nu_{4R} = C \bar \nu^T_{4L}$, with $C$ the charge conjugation matrix).
The interaction~\eqref{eq:interaction} could originate from the non-renormalizable, $SU(2)_L \times U(1)_Y$
invariant operator\footnote{In addition to the interaction~\eqref{eq:interaction}, this operator
can also induce $\bar \nu_{iR} \nu_{jL} \phi$ ($i, j = 1, 2, 3$), which mediate
unwanted invisible decays of the light neutrinos, $\nu_i \to \nu_j \phi$. To make sure that only
the interaction~\eqref{eq:interaction} is generated, it is enough, in the Dirac case, to assume the absence
of active-sterile mixing in the right-handed neutrino sector, such that $\nu_{sR} = \nu_{4R}$.
In the Majorana case, one must add the nonrenormalizable operators
$g_{\alpha \beta} (L_\alpha H) (L_\beta H) \phi / \Lambda$ ($\alpha, \beta = e, \mu, \tau$) with coefficients
$g_{\alpha e} = g_{se} U_{\alpha 4} / U^*_{s 4}$ (and $g_{\alpha \beta} = 0$ for $\alpha, \beta \neq e$),
such that only the interaction~\eqref{eq:interaction} arises at low energy~\cite{deGouvea:2019qre}.}
$g_{se}\, \bar \nu_{sR} L_e H \phi / \Lambda$, where $L_e$ is the lepton doublet
containing the left-handed electron and $\nu_{eL}$, $H$ is the Standard Model Higgs doublet,
and $\Lambda$ is a high scale associated with the underlying UV theory.

This scenario can be seen as a simplified model containing only the ingredients needed
to explain the LSND and MiniBooNE anomalies (namely, the mixing of the sterile neutrino
with the muon neutrino, and the decay of $\nu_4$ to the combination of light mass eigenstates
corresponding to the electon neutrino, $\nu_{eL} = \sum_{i=1}^3 U_{ei} \nu_{iL}$).
It differs on several points from the model of Ref.~\cite{Dentler:2019dhz}, in which
neutrinos are Dirac fermions, parity is conserved, both $U_{e4}$ and $U_{\mu4}$
are nonvanishing, $\nu_4$ can decay to all active neutrinos, and the light scalar field $\phi$
is unstable ($\phi \to \nu_i \bar \nu_j$, $i, j = 1, 2, 3$).
One could in principle generalize the model of Ref.~\cite{deGouvea:2019qre} by allowing
for a small mixing of the sterile neutrino with $\nu_e$ and $\nu_\tau$ (such that
$|U_{e4}|, |U_{\tau 4}| \ll |U_{\mu4}|$) and subdominant decay modes $\nu_4 \to \nu_\mu \phi$
and $\nu_4 \to \nu_\tau \phi$. This would not drastically change its phenomenology,
but in order to avoid introducing additional parameters with little impact
on the experimental signatures, we will stick
to the minimal assumptions~\eqref{eq:sterile_active_mixing} and~\eqref{eq:interaction}.

The HDSN scenario is constrained by several experimental measurements.
Searches for heavy sterile neutrinos in pion, kaon and muon decays constrain
$|U_{\mu 4}|^2 \lesssim 10^{-2}$ for $m_4 \gtrsim 1\, \mbox{MeV}$ (and $|U_{\mu 4}|^2 \ll 10^{-2}$
for $m_4$ between a few MeV and $\approx 400\, \mbox{MeV}$)~\cite{deGouvea:2015euy,Bryman:2019bjg}.
Below $m_4 \approx 1\, \mbox{MeV}$, the strongest constraint on $|U_{\mu 4}|^2$ comes from
$\nu_\mu$/$\bar \nu_\mu$ disappearance searches at MINOS/MINOS+, which give an upper bound
$|U_{\mu 4}|^2 < 2.3 \times 10^{-2}$ (90\% C.L.) for $m_4 \gtrsim 10\, \mbox{eV}$~\cite{MINOS:2017cae}.
The impact of $\nu_4$ decays on this bound is of higher order in $|U_{\mu 4}|^2$, as can be
seen from the survival probability~\eqref{eq:Pmumu}, and can be neglected.
The coupling $g$ of the $\nu_4 \nu_e \phi$ interaction is constrained by peak searches in leptonic meson
decays, giving $|g U_{\mu 4}|^2 < 1.9 \times 10^{-7}$ (90\% C.L.)~\cite{Pasquini:2015fjv},
which translates into $|g| \leq 2.8 \times 10^{-3}$ (respectively $|g| \leq 4.4 \times 10^{-2}$)
for $|U_{\mu 4}|^2$ equal to the MINOS upper bound (respectively $|U_{\mu 4}|^2 = 10^{-4}$).
We will consider a sterile neutrino mass in the range
$1\, \mbox{keV} \lesssim m_4 \lesssim 1\, \mbox{MeV}$ and values of $g$ such that
$0.01\, \mbox{eV} \leq |g| m_4 \leq 100\, \mbox{eV}$, consistent with the above experimental constraints.

Such a sterile neutrino could affect cosmological observations, and one has to make sure
that it is consistent with data. For the considered values of $g$ and $m_4$, all $\nu_4$'s
present in the early Universe decay well before recombination, leaving no impact
on the CMB and structure formation. It could instead affect Big Bang nucleosynthesis,
which requires the effective number of relativistic degrees of freedom $N_{\rm eff}$
to be close to 3 around $T = 1\, \mbox{MeV}$~\cite{Yeh:2022heq}.
However, this constraint can be evaded in the presence of\, ``secret interactions'' of the
sterile neutrino, e.g. by assuming that $\nu_s$ couples to an Abelian gauge boson~\cite{Hannestad:2013ana,Dasgupta:2013zpn}.
These interactions generate an effective, temperature-dependent potential for $\nu_s$
which suppresses its mixing with active neutrinos, thus preventing its production in
the early Universe until $T \ll 1\, \mbox{MeV}$~\cite{Dentler:2019dhz}.
Finally, the active-sterile mixing can induce $\phi$-mediated interactions between the light neutrino
mass eigenstates, in conflict with CMB observations, which require neutrinos to be free streaming
for redshifts $z \lesssim 10^5$~\cite{Cyr-Racine:2013jua,Forastieri:2017oma}.
However, such interactions do not arise in the scenario considered in this paper, in which
no $\nu_i \nu_j \phi$ couplings ($i, j = 1, 2,3$) are present. For the same reason, the SN 1987A constraint
on neutrino self-interactions~\cite{Kolb:1987qy} does not apply either.
%to this model.

Let us now discuss the short-baseline experimental signatures of the HDSN scenario.
We consider both the possibilities that neutrinos are Majorana or Dirac fermions.
In the Dirac case, the only allowed decay mode for the heavy sterile neutrino is $\nu_4 \to \nu_e \phi$
(and $\bar \nu_4 \to \bar \nu_e \phi$ for the antineutrino).
In the Majorana case, the 4-spinor $\nu_4$ satisfies the Majorana condition $\nu_4 = C \bar \nu^T_4$,
and the heavy neutrino can decay to both $\nu_e \phi$ and $\bar \nu_e \phi$ with equal decay rates.
%at tree level.
Given the energies of the LSND and MiniBooNE beams, the produced
$\nu_4$'s are relativistic, and their differential and total decay rates in the lab frame
(i.e., including an inverse Lorentz factor $1/\gamma_4 = m_4/E_4$)
are given by~\cite{Palomares-Ruiz:2005zbh}, in the Dirac case,
\begin{equation}
  \left. \frac{d \Gamma (\nu_4 \to \nu_e \phi)}{dE_e}\right|_{\rm Dirac}\, =\, \frac{|g|^2}{16 \pi}\, \frac{m^2_4 E_e}{E^3_4}\ ,
    \qquad  \Gamma^{\rm D}_4\, \equiv\,\left. \Gamma (\nu_4 \to \nu_e \phi)\right|_{\rm Dirac}\,
    =\, \frac{|g|^2 m^2_4}{32 \pi E_4}\ ,
\end{equation}
where $E_e \equiv E_{\nu_e}$ and $E_4 = E_{\nu_\mu}$ is the energy of $\nu_4$, which equals
the energy of the $\nu_\mu$ it originates from. The differential $\bar \nu_4$ decay rate is given
by the same expression. In the Majorana case, one has
\begin{equation}
 \left. \frac{d \Gamma (\nu_4 \to \nu_e \phi)}{dE_e}\right|_{\rm Majorana}\, =\,
    \left. \frac{d \Gamma (\nu_4 \to \bar \nu_e \phi)}{dE_e}\right|_{\rm Majorana}\, =\, \frac{|g|^2}{16 \pi}\, \frac{m^2_4 E_e}{E^3_4}\ ,
\end{equation}
\begin{equation}
  \Gamma^{\rm M}_4\, \equiv\,\left. \Gamma (\nu_4 \to \nu_e \phi)\right|_{\rm Majorana}
    + \left. \Gamma (\nu_4 \to \bar \nu_e \phi)\right|_{\rm Majorana}\, =\, \frac{|g|^2 m^2_4}{16 \pi E_4}\ .
\end{equation}

The probability for a $\nu_\mu$ to be observed as a $\nu_\mu$ or a $\nu_e$ (or a $\bar \nu_e$
in the Majorana case) after having travelled a distance $L$ depends on the fraction of $\nu_4$'s
in the $\nu_\mu$ beam, given by $|U_{\mu 4}|^2$, and on the $\nu_4$ decay rate into $\nu_e \phi$
($\bar \nu_e \phi$) in the lab frame.
Given the distance between the source and the detector in the LSND and MiniBooNE
experiments, standard oscillations between active flavours can safely be neglected\footnote{The
averaged active-sterile neutrino oscillations are taken into account in formulae~(\ref{eq:Pmumu}),
and have no impact on Eq.~(\ref{eq:Pmue_Dirac}), as $U_{e4} = 0$.}.
This also holds for the DUNE near detector considered in this paper.
The probabilities are then given by (denoting
$P_{\alpha \beta} \equiv P(\nu_\alpha \to \nu_\beta)$ and
$P_{\bar \alpha \bar \beta} \equiv P(\bar \nu_\alpha \to \bar \nu_\beta)$)~\cite{Palomares-Ruiz:2005zbh}:
\begin{align}
  P_{\mu \mu} = P_{\bar \mu \bar \mu} = (1 - |U_{\mu 4}|^2)^2 + |U_{\mu 4}|^4\, e^{-\Gamma_4 L}\, ,
    \qquad  P_{ee} = P_{\bar e \bar e} = 1\ ,\hspace{2cm}
\label{eq:Pmumu}
\end{align}
\begin{equation}
  P_{\mu e} = P_{\bar \mu \bar e} = |U_{\mu 4}|^2 (1 -  e^{-\Gamma_4 L})\ .\hspace{7cm}
\label{eq:Pmue_Dirac}
\end{equation}
The expressions~(\ref{eq:Pmumu}) for the survival probabilities are valid both in the Dirac and Majorana cases,
with the caveat that $\Gamma^{\rm M}_4 = 2\, \Gamma^{\rm D}_4$. For the appearance probabilities,
one has, in the Majorana case,
\begin{equation}
  P_{\mu e} = P_{\bar \mu \bar e} = P_{\mu \bar e} = P_{\bar \mu e} = \frac{1}{2}\, |U_{\mu 4}|^2 (1 -  e^{-\Gamma^{\rm M}_4 L})\ .
\label{eq:Pmue_Majorana}
\end{equation}
In our analysis, we will also need the dependence of the appearance probability $P_{\mu e}$
on the daughter electron neutrino energy, given by the differential probability
\begin{align}
  \frac{d P_{\mu e}}{d E_e} = \frac{d P_{\bar \mu \bar e}}{d E_e} = |U_{\mu 4}|^2 (1 -  e^{-\Gamma^{\rm D}_4 L})\, \frac{2 E_e}{E^2_4}  \qquad  &\text{(Dirac case)}\, ,  \\
  \frac{d P_{\mu e}}{d E_e} = \frac{d P_{\bar \mu \bar e}}{d E_e} = \frac{d P_{\mu \bar e}}{d E_e} = \frac{d P_{\bar \mu e}}{d E_e}
    = |U_{\mu 4}|^2 (1 -  e^{-\Gamma^{\rm M}_4 L})\, \frac{E_e}{E^2_4}   \qquad   &\text{(Majorana\ case)}\, .
\end{align}

In addition to its impact on the signal ($\nu_e$ appearance or $\nu_\mu$ disappearance), the heavy
sterile neutrino also affects the background. Let us consider for definiteness the neutrino mode.
The main backgrounds for the appearance signal ($\nu_e$ or $\bar \nu_e$ scattering
on Argon nuclei) are {\it (i)} the $\nu_e$/$\bar \nu_e$ contamination of the $\nu_\mu$ flux
(intrinsic background); {\it (ii)} the misidentification background (when the $\mu^-$/$\mu^+$ resulting
from the scattering of a $\nu_\mu$/$\bar \nu_\mu$ on a nucleus is misidentified as an electron/positron);
{\it (iii)} the neutral current background. While the intrinsic contamination background does
not depend on the physics scenario (Standard Model or heavy decaying sterile neutrino), the misidentification
background is proportional to the $\nu_\mu$ component of the incoming neutrino flux.
Since standard neutrino oscillations can be neglected,
the misidentification background in the HDSN scenario is suppressed by a factor $P_{\mu\mu}$
with respect to the Standard Model (SM):
\begin{equation}
  \frac{N^{\rm HDSN}_{\rm mis-ID}}{N^{\rm SM}_{\rm mis-ID}}\, = P_{\mu \mu}
    =  (1 - |U_{\mu 4}|^2)^2 + |U_{\mu 4}|^4\, e^{-\Gamma_4 L}\, ,
\label{eq:misID_bckgd}
\end{equation}
with $\Gamma_4 = \Gamma^{\rm D}_4$ in the Dirac case and $\Gamma_4 = \Gamma^{\rm M}_4$
in the Majorana case. Since $P_{\bar \mu \bar \mu} = P_{\mu\mu}$, Eq.~(\ref{eq:misID_bckgd}) holds
both in the neutrino and antineutrino modes.
As for the neutral current background, it is proportional to the fraction of active neutrinos
in the beam at the detector position, which is equal to $1$ in the SM,
but depleted by the sterile neutrino component in the HDSN scenario.
In the Dirac case, the suppression factor is given by $P_{\mu e} + P_{\mu \mu}$, yielding
\begin{equation}
  \frac{N^{\rm HDSN}_{\rm NC bckgd}}{N^{\rm SM}_{\rm NC bckgd}}\,
    = 1 - |U_{\mu 4}|^2 (1 - |U_{\mu 4}|^2) (1 +  e^{-\Gamma_4 L})  \qquad  {\rm (Dirac\ case})\ ,
\label{eq:NC_bckgd_Dirac}
\end{equation}
which is valid both in the neutrino and antineutrino modes, since
$P_{\bar \mu \bar e} + P_{\bar \mu \bar \mu} = P_{\mu e} + P_{\mu \mu}$.
In the Majorana case, the muon (anti)neutrinos
can decay to both $\nu_e$ and $\bar \nu_e$, which have different neutral current cross sections
$\sigma^{\rm NC}_{\nu_e}$ and $\sigma^{\rm NC}_{\bar \nu_e}$. One therefore has, in the neutrino mode,
\begin{eqnarray}
  \frac{N^{\rm HDSN}_{\rm NC bckgd}}{N^{\rm SM}_{\rm NC bckgd}}\
    & =\, & P_{\mu \mu} + P_{\mu e} + P_{\mu \bar e}\, \frac{\sigma^{\rm NC}_{\bar \nu_e}}{\sigma^{\rm NC}_{\nu_e}}
    \qquad  {\rm (Majorana\ case,\, neutrino\ mode})  \nonumber   \\
    & \approx\, & 1 - \frac{5}{4}\, |U_{\mu 4}|^2 \left( 1 + \frac{3}{5}\, e^{-\Gamma^M_4 L} \right)
      + |U_{\mu 4}|^4 (1 +  e^{-\Gamma^M_4 L}) \ ,
\label{eq:NC_bckgd_Majorana_nu}
\end{eqnarray}
where we neglected the energy dependence of the ratio $\sigma^{\rm NC}_{\bar \nu_e} / \sigma^{\rm NC}_{\nu_e}$,
and we used $\sigma^{\rm NC}_{\nu_e} \approx 2 \sigma^{\rm NC}_{\bar \nu_e}$
and $\sigma^{\rm NC}_{\nu_\mu} = \sigma^{\rm NC}_{\nu_e}$, while in the antineutrino mode
\begin{eqnarray}
  \frac{N^{\rm HDSN}_{\rm NC bckgd}}{N^{\rm SM}_{\rm NC bckgd}}\
    & =\, & P_{\bar \mu \bar \mu} + P_{\bar \mu \bar e} + P_{\bar \mu e}\, \frac{\sigma^{\rm NC}_{\nu_e}}{\sigma^{\rm NC}_{\bar \nu_e}}
    \qquad  {\rm (Majorana\ case,\, antineutrino\ mode})  \nonumber  \\
    & \approx\, & 1 - \frac{1}{2}\, |U_{\mu 4}|^2 \left( 1 + 3\, e^{-\Gamma^M_4 L} \right)
      + |U_{\mu 4}|^4 (1 +  e^{-\Gamma^M_4 L}) \ .
\label{eq:NC_bckgd_Majorana_antinu}
\end{eqnarray}
Note that, in the limit $e^{-\Gamma_4 L} \to 1$ (which corresponds to the case where only
a negligible fraction of the sterile neutrinos decay before reaching the detector),
the right-hand sides of Eq.~(\ref{eq:misID_bckgd}) to~(\ref{eq:NC_bckgd_Majorana_antinu})
all reduce to $1 - 2 |U_{\mu 4}|^2 (1 - |U_{\mu 4}|^2)$.

Finally, the main background for the disappearance signal ($\nu_\mu$ or $\bar \nu_\mu$ scattering
on Argon nuclei) is the neutral current background, given by Eqs.~(\ref{eq:NC_bckgd_Dirac})
to~(\ref{eq:NC_bckgd_Majorana_antinu}).

%%%%%%%%%%%%%%
\section{Experimental setup and simulation details}
\label{sec:experiment}
%%%%%%%%%%%%%%

In this section, we briefly discuss the experimental specifications of the DUNE far and near detector facilities,
before describing in detail the simulations performed in our analysis. 

\subsection{Experimental setup}
\label{sec:Detector}

The next generation of long-baseline oscillation experiments will dedicate their main efforts to searching for the Dirac CP-violating phase of the lepton mixing matrix, a fundamental parameter of the Standard Model, and a key observable for a better understanding of the Universe in its first stages.  One of these experiments will be the Deep
Underground Neutrino Experiment (DUNE) at Fermilab. According to the Technical Design Report configuration~\cite{DUNE:2020jqi}, DUNE will use a 120 GeV proton beam with a power of 1.2 MW, yielding $1.1 \times 10^{21}$ protons on target per year. This experiment will have two sets of detectors. The far detector, based on a liquid argon time projection chamber,
will be placed approximately 1.5~km underground and 1300~km downstream of the source in the Stanford Underground Research
Facility (SURF) in South Dakota. It will have a total mass of 70 kt and a fiducial mass of roughly 40 kt. The near
detector complex will be located at Fermilab, approximately 574~m away from the neutrino source. It will consist of three
different detectors, two of which will be able to move and collect data at different angles from the on-axis position.

\begin{figure}
\centering
\includegraphics[height=6.7cm, width=6.7cm]{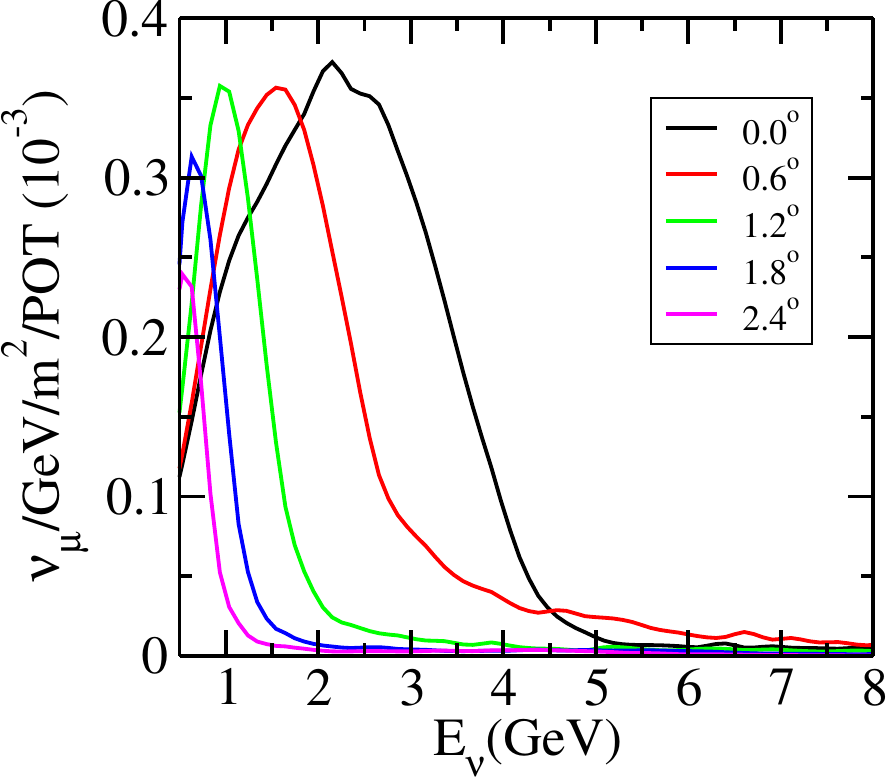}
\hskip 1cm
\includegraphics[height=6.7cm, width=6.7cm]{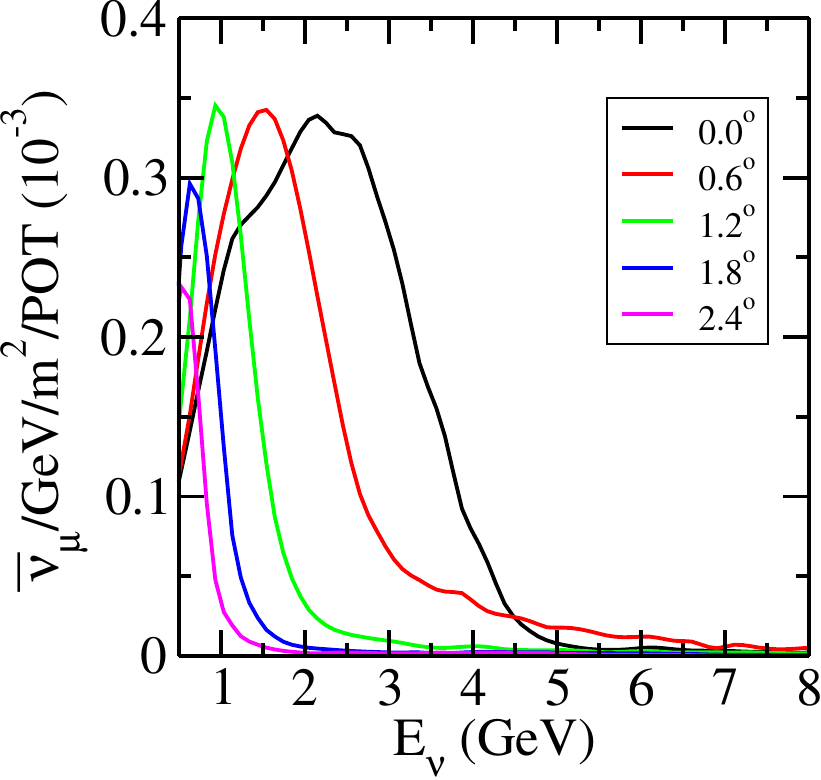}
\caption{Expected muon neutrino (left) and antineutrino (right) fluxes at the movable DUNE liquid argon
near detector (ND-LAr), for on-axis and several off-axis locations.}
\label{fig:nuSpec}
\end{figure}

In this work, we consider the $67.2\,\rm t$ movable liquid argon near detector (ND-LAr)
proposed by the DUNE Collaboration~\cite{DUNE:2021tad}.
Fig.~\ref{fig:nuSpec} represents the muon (anti)neutrino flux as a function of energy for on-axis and different off-axis
positions of the detector~\cite{DeRomeri:2019kic,DUNE:2021tad}. As can be seen from the figure,
the (anti)neutrino flux becomes smaller and the energy spectrum shifts towards lower energies as the off-axis angle increases.

%%%%%%%%%%%%%%%%
\subsection{Simulation details}
%%%%%%%%%%%%%%%%

Let us now discuss how the appearance and disappearance event spectra at the DUNE liquid argon near detector
are calculated, and how the ability of the detector
to test the heavy decaying sterile neutrino hypothesis is assessed.
In order to estimate the statistical sensitivity of ND-LAr to the HDSN parameters, we assume the standard
3-neutrino framework as the true hypothesis, and we use the Poissonian $\chi^2$ function
\begin{equation}
  \chi^2(\kappa)\; =\; \underset{\xi}{min}\left[\sum\limits_{i}2\left\{N_i(\kappa, \xi) - N_{i}^{obs}
    + N_{i}^{obs}\,{\rm ln}\left(\frac{N_{i}^{obs}}{N_i(\kappa, \xi)}\right) \right\} + \sum\limits_{\xi}\left(\frac{\xi}{\sigma_{\xi}}\right)^2  \right],
\label{eq:analysis}
\end{equation}
where $i$ labels the different energy bins, $N_i^{obs}$ is the number of events in the $i$th bin predicted by the SM,
and $N_i(\kappa, \xi)$ is the corresponding quantity in the presence of the heavy decaying sterile neutrino.
The latter depends on the HDSN parameters, collectively denoted by $\kappa$, and on normalization factors
for the signal and background, $\xi_{sig}$ and $\xi_{bg}$. Explicitly,
\begin{equation}
  N_i^{obs} = N^{sig}_i (SM) + N^{bg}_i (SM)\; ,  \qquad
    N_i(\kappa, \xi) = N^{sig}_i (\kappa) (1+ \xi_{sig}) + N^{bg}_i (\kappa) (1 + \xi_{bg})\, ,
\label{eq:Ni}
\end{equation}
where $N^{sig}_i (SM)$ and $N^{sig}_i (\kappa)$ are the numbers of signal events in the SM and in the HDSN scenario, respectively, and similarly for the background events $N^{bg}_i (SM)$ and $N^{bg}_i (\kappa)$. The normalization factors $\xi_{sig}$ and $\xi_{bg}$ quantify the systematic errors, with associated uncertainties $\sigma_{sig}$ and $\sigma_{bg}$. We assume these systematic uncertainties to be uncorrelated, and conservatively take $\sigma_{sig} = \sigma_{bg} = 10\%$ accross all energy bins\footnote{We are assuming energy-independent systematic uncertainties in this analysis, due to the lack of publicly available information about their energy dependence.}.

Let us consider the number of events predicted by the SM in a given signal or background channel.
Assuming this channel involves a flavour transition $\nu_\alpha \to \nu_\beta$ with probability
$P_{\alpha \beta}$ (with $\alpha = \beta$ or $\alpha \neq \beta$), the number of events
in the $i$th reconstructed energy bin $[E_{rec}^i, E_{rec}^{i+1}]$ is given by the following formula:
\begin{equation}
  N^{\nu_\alpha \to \nu_\beta}_i = \frac{nT}{4\pi L^2}\, \int_{E_{rec}^i}^{E_{rec}^{i+1}}\!\! A(E_{rec})\,dE_{rec}
    \int_{0}^{{E}_{\nu}^{max}} dE_{\nu}\,\mathcal{R}(E_{\nu},E_{rec})\,\sigma_{\nu_\beta}(E_{\nu})\,
    \phi_{\nu_\alpha}(E_{\nu})\, P_{\alpha\beta}\, ,
\label{SM_Events}
\end{equation}
where $E_\nu$ and $E_{rec}$ are the true and reconstructed neutrino energies, respectively,
$\phi_{\nu_\alpha}(E_{\nu})$ is the flux of the $\nu_{\alpha}$ component of the neutrino beam,
$\sigma_{\nu_\beta}(E_{\nu})$ the interaction cross-section of $\nu_{\beta}$,
$\mathcal{R}(E_{\nu},E_{rec})$ the energy resolution function of the detector, and $A(E_{rec})$
the detector efficiency.
The upper limit ${E}_{\nu}^{max}$ of the second integral corresponds to the maximal energy of the muon (anti)neutrinos
in the beam, and the prefactor involves the total running time $T$, the number $n$ of target Argon nuclei
in the detector, and the distance $L$ between the source and the detector.
All details regarding the cross sections, energy resolution and efficiency of the detector
that we use in our simulations can be found in Ref.~\cite{DUNE:2021cuw}.

\begin{table}[t]
{
\newcommand{\mc}[3]{\multicolumn{#1}{#2}{#3}}
\newcommand{\mr}[3]{\multirow{#1}{#2}{#3}}
\centering
\begin{adjustbox}{width=1\textwidth}
 %\small
\begin{tabular}{|c||c|c|c|c|}
\hline
\mr{3}{*}{Channel} & \mc{4}{c|}{ND-LAr event rates ($\times 10^6$)} \\
\cline{2-5}
      & Signal (CC) & \mc{3}{c|}{Background}  \\
\cline{2-5}
      &  {Right (wrong) sign component} & Intrinsic & Mis-ID & NC \\
\hline
$\nu$ mode: $\nu_{\mu}\rightarrow\nu_{e}$ & $\simeq 0\ (\simeq 0)$ & 2.14 & 0.43 & 0.65\\

\hline
$\bar{\nu}$ mode: $\rm{\bar{\nu}_{\mu}\rightarrow \bar{\nu}_{e}}$ & $\simeq 0\ (\simeq 0)$ & 1.2 & 0.11  & 0.38\\

\hline
$\nu$ mode: $\nu_{\mu}\rightarrow\nu_{\mu}$ &  206 (7.0) & $-$ & $-$ & 1.71 \\

\hline
$\bar{\nu}$ mode: $\rm{\bar{\nu}_{\mu}\rightarrow \bar{\nu}_{\mu}}$ &  78 (20) & $-$ & $-$ & 0.98\\

\hline
\end{tabular}
\end{adjustbox}
}
\caption{Event rates at the DUNE liquid argon near detector (ND-LAr) expected in the Standard Model,
assuming 3.5 years of on-axis data taking in either neutrino or antineutrino mode.
The first and second rows correspond to the appearance channel
in the neutrino and antineutrino modes, respectively, for which the signal from neutrino oscillations is negligible.
The last three columns show the number of intrinsic, misidentification, and neutral current background events,
as defined in Section~\ref{sec:NuDecay}.
The third and fourth rows correspond to the disappearance channel in the neutrino and antineutrino modes,
respectively, and the number of signal events is given separately for the right sign component
(e.g. $\nu_{\mu} \rightarrow \nu_{\mu}$ in the neutrino mode) and for the wrong sign component
(e.g. $\bar \nu_{\mu} \rightarrow \bar \nu_{\mu}$ in the neutrino mode) of the neutrino beam.}
\label{table1}
\end{table}

Table~\ref{table1} displays the expected numbers of signal and background events in the appearance
and disappearance channels at the DUNE ND-LAr detector, both in the neutrino and antineutrino modes,
assuming 3.5 years of on-axis data taking in each mode. For the appearance channel, the numbers of
intrinsic, misidentification and neutral current background events are given separately.
The negligible appearance signal is due to the fact that neutrino oscillations do not
have the time to develop significantly over such a short baseline.

In the HDSN scenario, Eq.~(\ref{SM_Events}) must be modified to take into account
the sterile neutrino decays. Let us first consider the case where neutrinos are Dirac fermions
and the experiment runs in the neutrino mode. The number of appearance signal events
in the $i$th energy bin is given by Eq.~(\ref{eq:events_nu_appearance_Dirac}), where
the last integral accounts for the fact that the daughter $\nu_e$ is less energetic than
the parent $\nu_\mu$. Eqs.~(\ref{eq:events_nu_disappearance_Dirac}) and~(\ref{eq:events_nu_NC_Dirac})
correspond to the disappearance signal and to the neutral current background, respectively.
\begin{align}
\left[N^{\nu_{\mu}\rightarrow \nu_e}_{i}\right]^{D,\, CC}  & = \frac{nT}{4\pi L^2}\,\int_{E_{rec}^i}^{E_{rec}^{i+1}}\!\! A(E_{rec})\, dE_{rec}  \int_{0}^{E_{\nu}^{max}}\!\!\! dE_{\nu_e}\,\mathcal{R}(E_{\nu_e},\,E_{rec})\,\sigma^{CC}_{\nu_{e}}(E_{\nu_e})  \nonumber \hspace{1.3cm} \\
&\times \int_{E_{\nu_e}}^{E_{\nu}^{max}}\!\!\! dE_{\nu_{\mu}}\, \phi_{\nu_{\mu}}(E_{\nu_{\mu}})\,
\frac{dP^D_{\mu e}}{dE_{\nu_e}}\, (E_{\nu_{\mu}}, E_{\nu_e})\, ,
\label{eq:events_nu_appearance_Dirac}
\end{align}
\begin{align}
\left[N^{\nu_{\mu}\rightarrow \nu_\mu}_{i}\right]^{D,\, CC}  & = \frac{nT}{4\pi L^2}\,\int_{E_{rec}^i}^{E_{rec}^{i+1}}\!\! A(E_{rec})\,dE_{rec} \nonumber \hspace{7.1cm}\\
&\times \int_{0}^{E_{\nu}^{max}}\!\!\! dE_{\nu_{\mu}}\,\mathcal{R}(E_{\nu_{\mu}},\,E_{rec})\, \sigma^{CC}_{\nu_{\mu}}(E_{\nu_{\mu}})\, \phi_{\nu_{\mu}}(E_{\nu_{\mu}})\, P^D_{\mu \mu}(E_{\nu_{\mu}})\, ,
\label{eq:events_nu_disappearance_Dirac}
\end{align}
\begin{align}
\left[N^{\nu_\mu}_{i}\right]&^{D,\,NC}  = \frac{nT}{4\pi L^2}\,\int_{E_{rec}^i}^{E_{rec}^{i+1}}\!\! A(E_{rec})\, dE_{rec} \nonumber
\hspace{7.7cm}\\
&\times \biggl[\int_{0}^{E_{\nu}^{max}}\!\!\! dE_{\nu_{\mu}}\,\mathcal{R}(E_{\nu_{\mu}},E_{rec})\, \sigma^{NC}_{\nu}(E_{\nu_\mu})\, \phi_{\nu_{\mu}}(E_{\nu_{\mu}})\, P^D_{\mu \mu}(E_{\nu_{\mu}})  \nonumber \\
&+ \int_{0}^{E_{\nu}^{max}}\!\!\! dE_{\nu_e}\,\mathcal{R}(E_{\nu_e},E_{rec})\,\sigma^{NC}_{\nu}(E_{\nu_e}) \int_{E_{\nu_e}}^{E_{\nu}^{max}}\!\!\! dE_{\nu_{\mu}}\, \phi_{\nu_{\mu}}(E_{\nu_{\mu}})\, \frac{dP^D_{\mu e}}{dE_{\nu_e}}\, (E_{\nu_{\mu}}, E_{\nu_e}) \biggr]\, .
\label{eq:events_nu_NC_Dirac}
\end{align}
In the Majorana case, the sterile neutrino can decay either to a $\nu_e$ or to a $\bar \nu_e$,
and Eqs.~(\ref{eq:events_nu_appearance_Dirac}) to~\eqref{eq:events_nu_NC_Dirac} are replaced
by the following equations:
\begin{align}
\left[N^{\nu_{\mu}\rightarrow \nu_e/\bar{\nu}_e}_{i}\right]^{M,\,CC}  &= \frac{nT}{4\pi L^2}\,\int_{E_{rec}^i}^{E_{rec}^{i+1}}\!\! A(E_{rec})\,dE_{rec}\ \biggl[\int_{0}^{E_{\nu}^{max}}\!\!\! dE_{\nu_e}\,\mathcal{R}(E_{\nu_e},E_{rec})\,\sigma^{CC}_{\nu_{e}}(E_{\nu_e})  \nonumber
\\
%\hspace{14cm} \\
\times \int_{E_{\nu_e}}^{E_{\nu}^{max}}\!\!\! dE&_{\nu_{\mu}}\, \phi_{\nu_{\mu}}(E_{\nu_{\mu}})\, \frac{dP^M_{\mu e}}{dE_{\nu_e}}\, (E_{\nu_{\mu}}, E_{\nu_e}) \nonumber\\
+ \int_{0}^{E_{\nu}^{max}}\!\!\! dE&_{\bar{\nu}_e}\, \mathcal{R}(E_{\bar{\nu}_e},E_{rec})\,\sigma^{CC}_{\bar{\nu}_{e}}(E_{\bar{\nu}_e})
%\nonumber \\
%& \times
\int_{E_{\bar{\nu}_e}}^{E_{\nu}^{max}}\!\!\! dE_{\nu_{\mu}}\, \phi_{\nu_{\mu}}(E_{\nu_{\mu}})\, \frac{dP^M_{\mu \bar{e}}}{dE_{\bar{\nu}_e}}\, (E_{\nu_{\mu}}, E_{\bar{\nu}_e}) \biggr]\, ,
\label{eq:events_nu_appearance_Majorana}
\end{align}
\begin{align}
\left[N^{\nu_{\mu}\rightarrow \nu_{\mu}}_{i}\right]^{M,\,CC} & = \frac{nT}{4\pi L^2}\,\int_{E_{rec}^i}^{E_{rec}^{i+1}}\!\! A(E_{rec})\,dE_{rec} \nonumber \hspace{7.1cm} \\
&\times \int_{0}^{E_{\nu}^{max}}\!\!\! dE_{\nu_{\mu}}\,\mathcal{R}(E_{\nu_{\mu}},E_{rec})\, \sigma^{CC}_{\nu_{\mu}}(E_{\nu_{\mu}})\, \phi_{\nu_{\mu}}(E_{\nu_{\mu}})\, P^M_{\mu \mu}(E_{\nu_{\mu}})\, ,
\label{eq:events_nu_disappearance_Majorana}
\end{align}
\begin{align}
\left[N^{\nu_\mu}_{i}\right]&^{M,\,NC}  = \frac{nT}{4\pi L^2}\,\int_{E_{rec}^i}^{E_{rec}^{i+1}}\!\! A(E_{rec})\, dE_{rec}
\nonumber \\
%\hspace{10cm}  \\ 
\times& \, \biggl[\int_{0}^{E_{\nu}^{max}}\!\!\! dE_{\nu_{\mu}}\,\mathcal{R}(E_{\nu_{\mu}},E_{rec})\, \sigma^{NC}_{\nu}(E_{\nu_\mu})\, \phi_{\nu_{\mu}}(E_{\nu_{\mu}})\, P^M_{\mu \mu}(E_{\nu_{\mu}})  \nonumber \\
+& \int_{0}^{E_{\nu}^{max}}\!\!\! dE_{\bar{\nu}_e}\,\mathcal{R}(E_{\bar{\nu}_e},E_{rec})\,\sigma^{NC}_{\bar{\nu}}(E_{\bar{\nu}_e})\, \int_{E_{\bar{\nu}_e}}^{E_{\nu}^{max}}\!\!\! dE_{\nu_{\mu}}\, \phi_{\nu_{\mu}}(E_{\nu_{\mu}})\, \frac{dP^M_{\mu \bar{e}}}{dE_{\bar{\nu}_e}}\, (E_{\nu_{\mu}}, E_{\bar{\nu}_e}) \nonumber \\
+& \int_{0}^{E_{\nu}^{max}}\!\!\! dE_{\nu_e}\,\mathcal{R}(E_{\nu_e},E_{rec})\,\sigma^{NC}_{\nu}(E_{\nu_e}) \int_{E_{\nu_e}}^{E_{\nu}^{max}}\!\!\! dE_{\nu_{\mu}}\, \phi_{\nu_{\mu}}(E_{\nu_{\mu}})\, \frac{dP^M_{\mu e}}{dE_{\nu_e}}\, (E_{\nu_{\mu}}, E_{\nu_e}) \biggr)\biggr]\, .
\label{eq:events_nu_NC_Majorana}
\end{align}

It is straightforward to generalize Eqs.~(\ref{eq:events_nu_appearance_Dirac}) to~\eqref{eq:events_nu_NC_Majorana}
to the case where the experiment runs in the antineutrino mode.
The probabilities appearing in these equations are given by (see Section~\ref{sec:NuDecay})
\begin{align}
\frac{d P^D_{\mu e}}{d E_e}  = \frac{d P^D_{\bar \mu \bar e}}{d E_e} = |U_{\mu 4}|^2 (1 -  e^{-\Gamma^{\rm D}_4 L})\, \frac{2 E_e}{E^2_4}  \qquad  &\text{(Dirac case})\, , \\
\frac{d P^M_{\mu e}}{d E_e} = \frac{d P^M_{\bar \mu \bar e}}{d E_e} = \frac{d P^M_{\mu \bar e}}{d E_e} = \frac{d P^M_{\bar \mu e}}{d E_e}
    = |U_{\mu 4}|^2 (1 - e^{-\Gamma^{\rm M}_4 L})\, \frac{E_e}{E^2_4}  \qquad  &\text{(Majorana case})\, ,
\end{align}
where, depending on the case, $E_e = E_{\nu_e}$ or $E_{\bar \nu_e}$, $E_4 = E_{\nu_\mu}$ or $E_{\bar \nu_\mu}$, and
\begin{align}
 & P^D_{\mu \mu} = P^D_{\bar \mu \bar \mu} = \left(1-|U_{\mu 4}|^2\right)^2 + |U_{\mu 4}|^4  e^{-\Gamma^{\rm D}_4 L} \qquad \text{(Dirac case)}\, , \hspace{2cm}  \\
 & P^M_{\mu \mu} = P^M_{\bar \mu \bar \mu} = \left(1-|U_{\mu 4}|^2 \right)^2 + |U_{\mu 4}|^4  e^{-\Gamma^{\rm M}_4 L}  \qquad \text{(Majorana case)}\, . 
\end{align}

%%%%%%%%%%%%%%%%%%%%%%%%%
\section{ND-LA${\textrm r}$ sensitivity to the HDSN parameters}
\label{sec:sensitivity}
%%%%%%%%%%%%%%%%%%%%%%%%%

In this section, we investigate the possibility to test the heavy decaying sterile neutrino (HDSN)
scenario at the DUNE liquid argon near detector (ND-LAr).
As we are going to see, ND-LAr has the capability to exclude a large portion of the HDSN parameter space,
much bigger than the one consistent with the LSND and MiniBooNE anomalies, which is currently
probed by the MicroBooNE experiment at Fermilab.

%%%%%%%%%%%%%%%%%%%%%%%%%%%%%%%%%%%%%%
\subsection{Appearance and disappearance spectra in the Standard Model}
%%%%%%%%%%%%%%%%%%%%%%%%%%%%%%%%%%%%%%
\begin{figure}
\centering
\includegraphics[height=7cm, width=7cm]{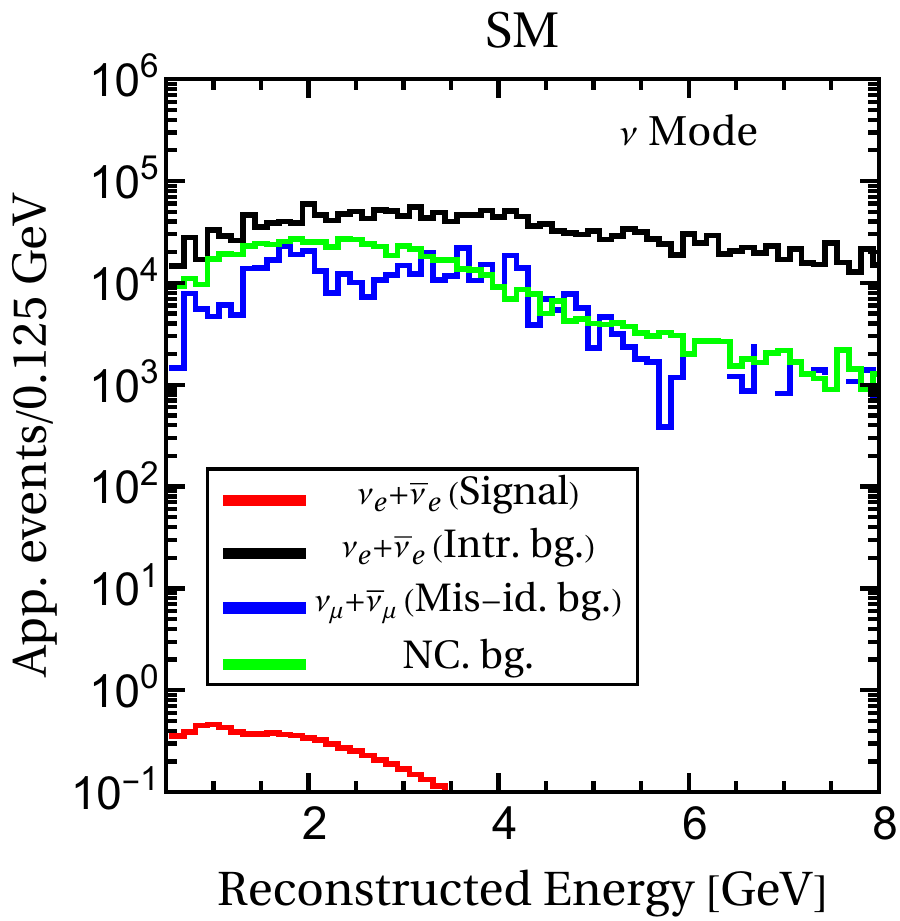}
\hskip .8cm
\includegraphics[height=7cm, width=7cm]{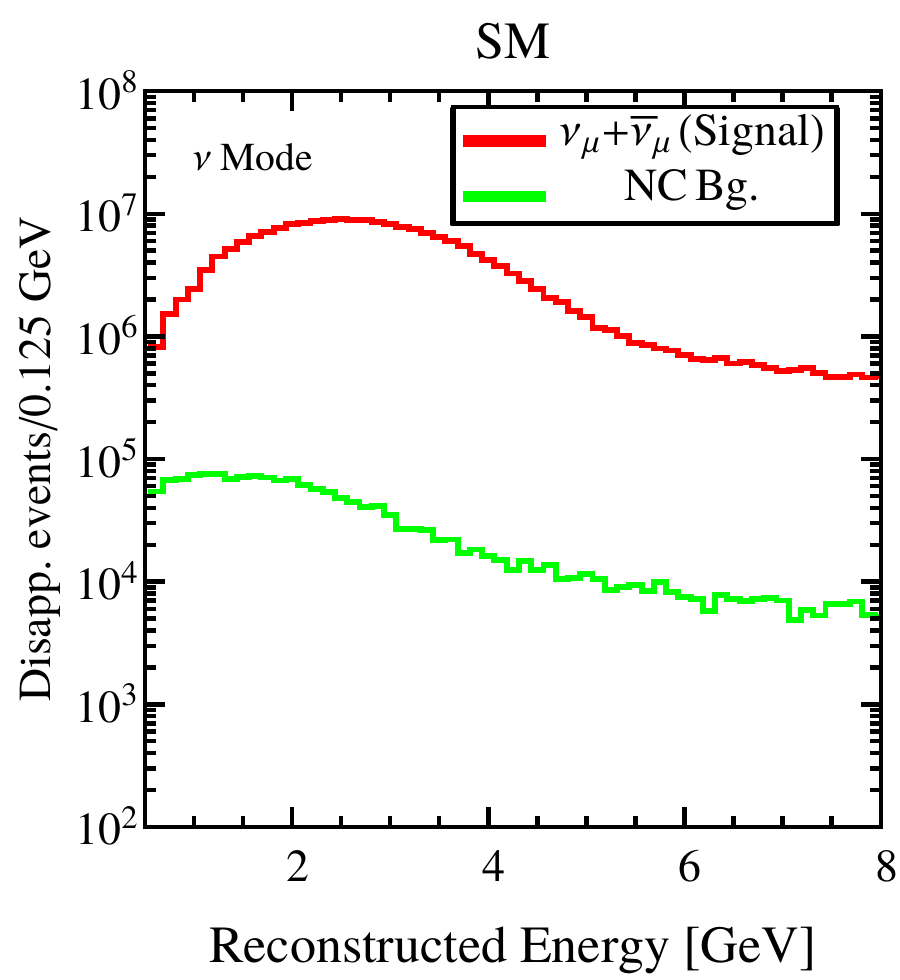}\\
\vskip .3cm
\includegraphics[height=7cm, width=7cm]{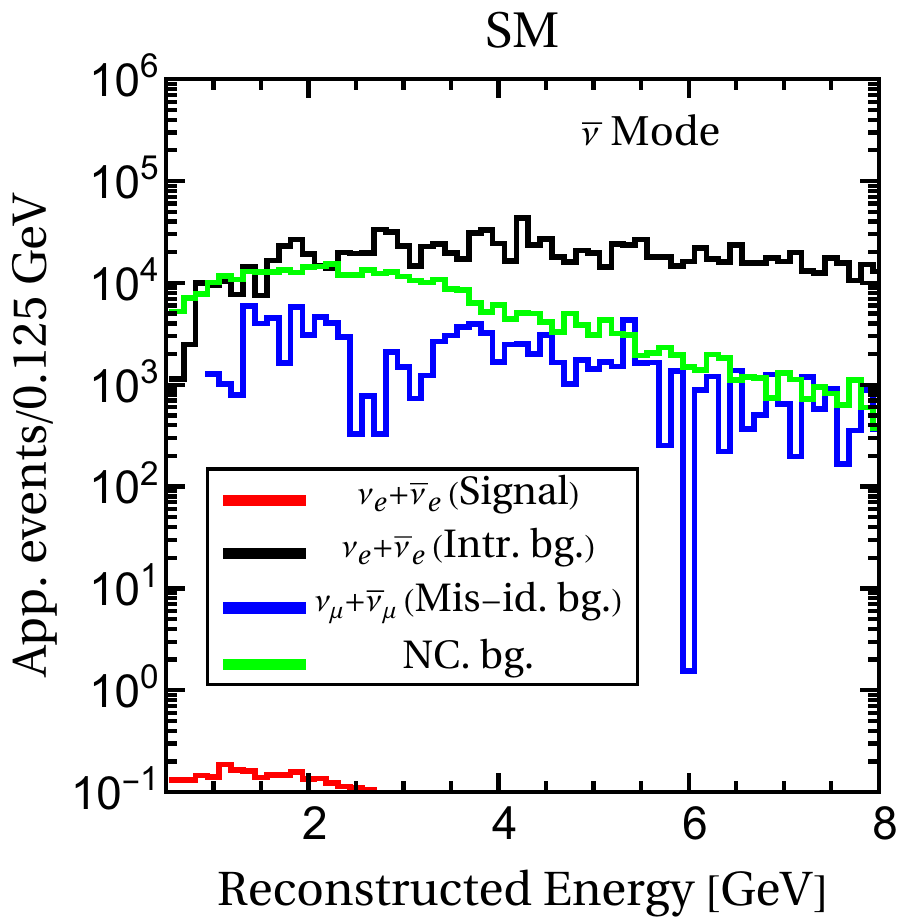}
\hskip .8cm
\includegraphics[height=7cm, width=7cm]{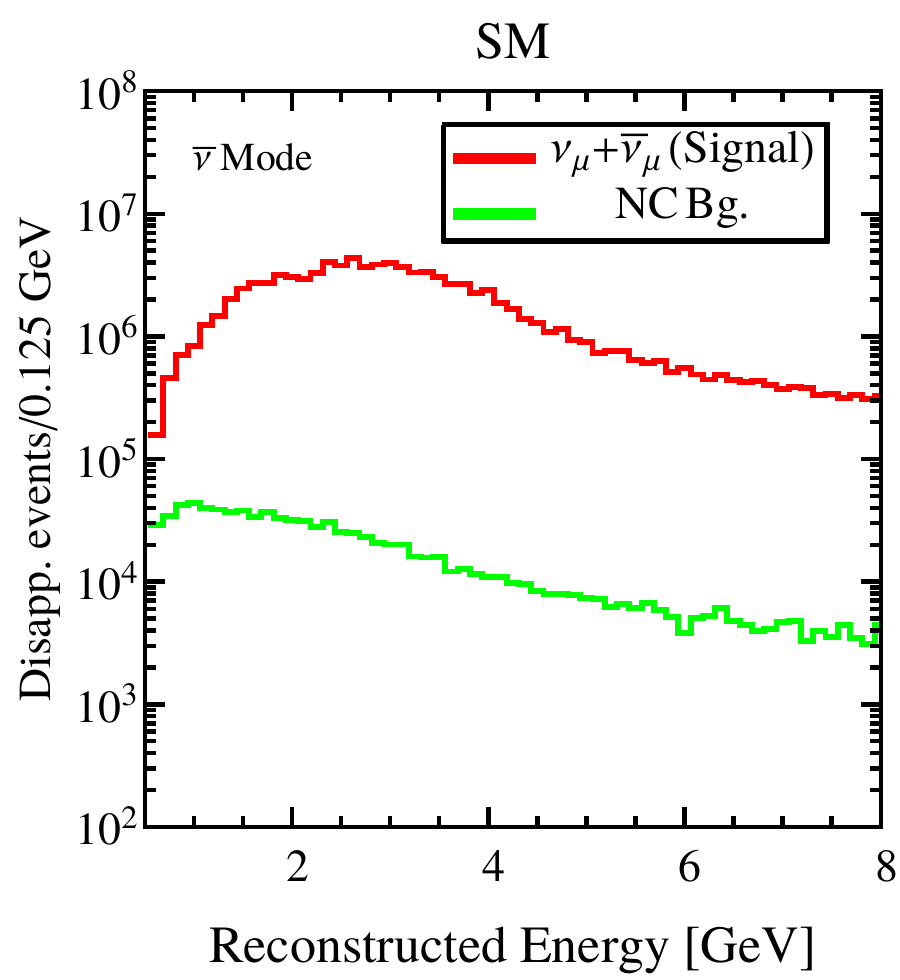}
\caption{Appearance (left) and disappearance (right) event spectra expected at the DUNE ND-LAr detector
in the Standard Model, assuming a running time of 3.5 years on axis in either neutrino
or antineutrino mode. The top and bottom panels correspond to the neutrino and antineutrino modes, respectively.
In each plot, the signal and the various background events are represented with different colours (see legend).}
\label{fig:ND_spectra_SM}
\end{figure}
Let us first consider the charged current signal from neutrino-nucleus scattering
and the associated backgrounds. We are interested in both the appearance
($\nu_\mu \to \nu_e$ and $\bar \nu_\mu \to \bar \nu_e$) and disappearance
($\nu_\mu \to \nu_\mu$ and $\bar \nu_\mu \to \bar \nu_\mu$) channels.
Since the ND-LAr detector is not magnetized,
we assume that muons are not distinguished from antimuons,
such that both $\nu_\mu$'s and $\bar \nu_\mu$'s from the neutrino beam contribute to the
disappearance signal, irrespective of whether the experiment runs in the neutrino
or antineutrino mode. A similar statement holds for the appearance channel (with the additional
subtlety, in the HDSN scenario, that $\nu_\mu \to \bar \nu_e$ and $\bar \nu_\mu \to \nu_e$
transitions are also possible if neutrinos are Majorana fermions).

Fig.~\ref{fig:ND_spectra_SM} shows the (anti)neutrino spectra (i.e., the number of appearance or disappearance
events as a function of the reconstructed (anti)neutrino energy) expected at the DUNE ND-LAr detector
in the Standard Model, assuming 3.5 years of running time both in the neutrino and antineutrino
modes. The left and right panels display the appearance and disappearance spectra,
respectively, while the top and bottom panels correspond to the neutrino and antineutrino modes.
In each plot, the signal and the various background events are represented with different colours.
Appearance events (red curve in the left panels) refer to electrons or positrons
produced from the scattering of a $\nu_e$ or a $\bar \nu_e$ on an Argon nucleus. The background events
that can mimick such a signal are the intrisic background from the $\nu_e$ and $\bar \nu_e$
contamination of the neutrino flux (black); the misidentification backgound due to negative or positive muons
misidentified as electrons or positrons (blue);
and the neutral current background (green), to which all active neutrinos and antineutrinos contribute.
As expected, the SM signal in the appearance channel is extremely small, due
to the short baseline of the near detector, which prevents $\nu_\mu \to \nu_e$ and
$\bar \nu_\mu \to \bar \nu_e$ oscillations to develop. The observed events are almost exclusively
background events.
In the disappearance channel (right panels), signal events correspond to muons or antimuons
produced from the scattering of a $\nu_\mu$ or a $\bar \nu_\mu$ on an Argon nucleus.
As can be seen from the plots, the SM signal (red) strongly dominates over the neutral current
background (green).
\begin{figure}
\centering
\includegraphics[height=4.7cm, width=4.7cm]{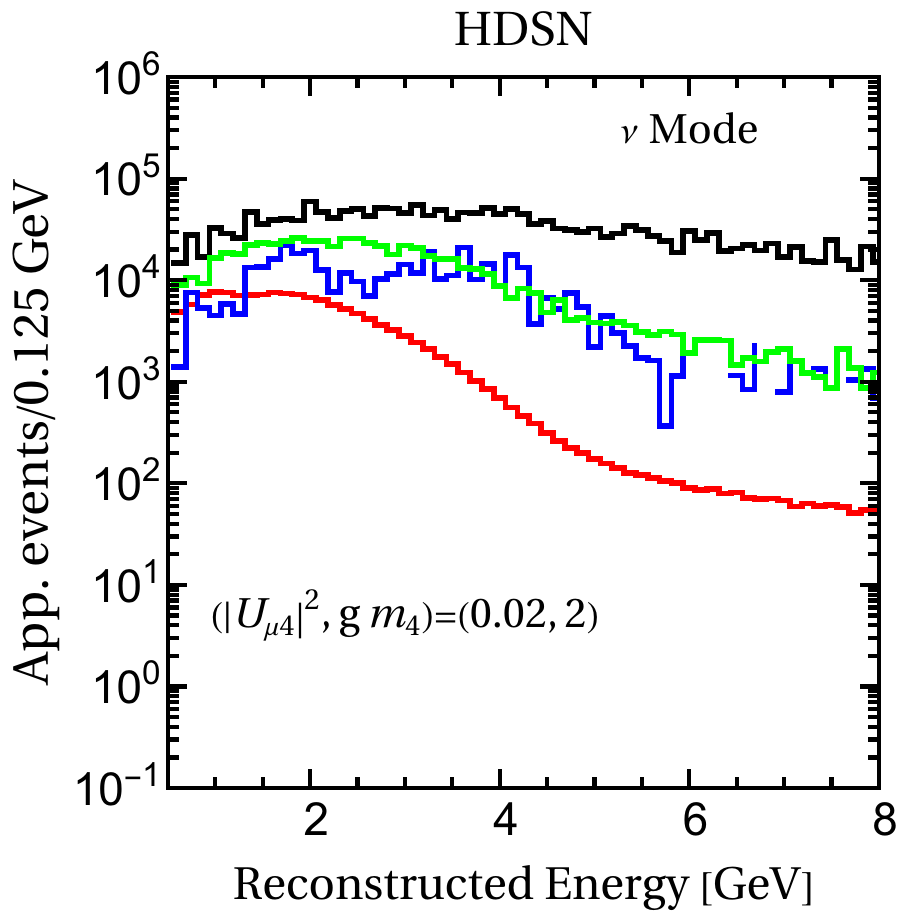}
\hskip .3cm
\includegraphics[height=4.7cm, width=4.7cm]{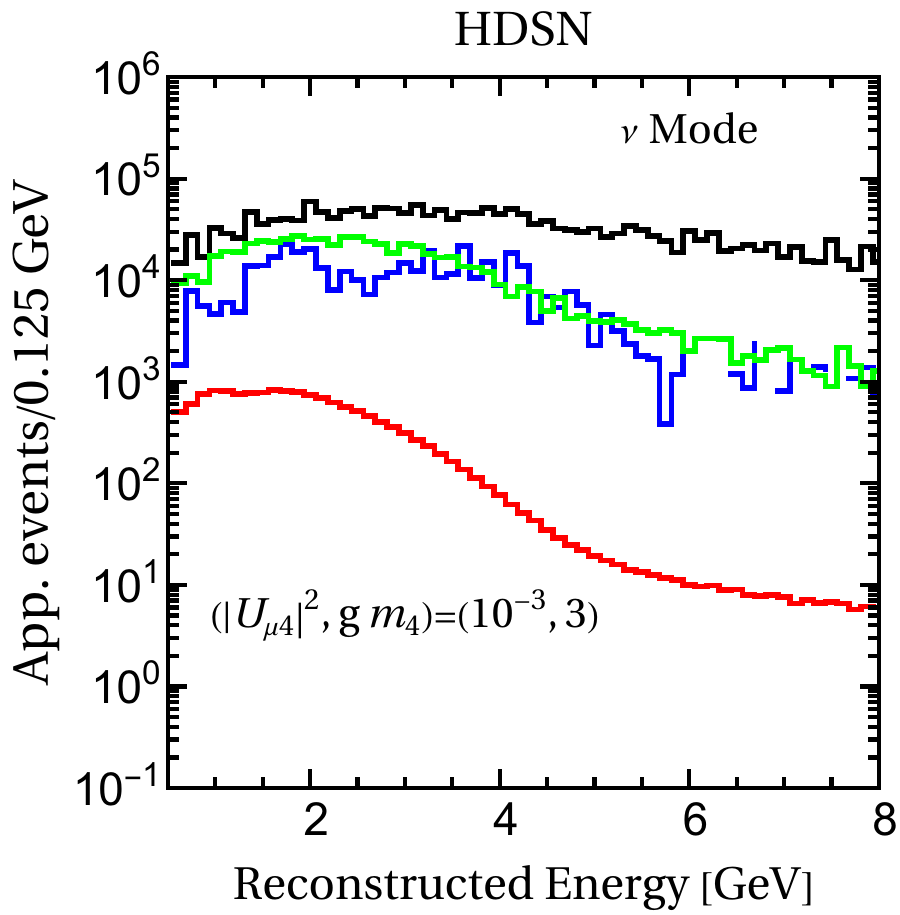}
\hskip .3cm
\includegraphics[height=4.7cm, width=4.7cm]{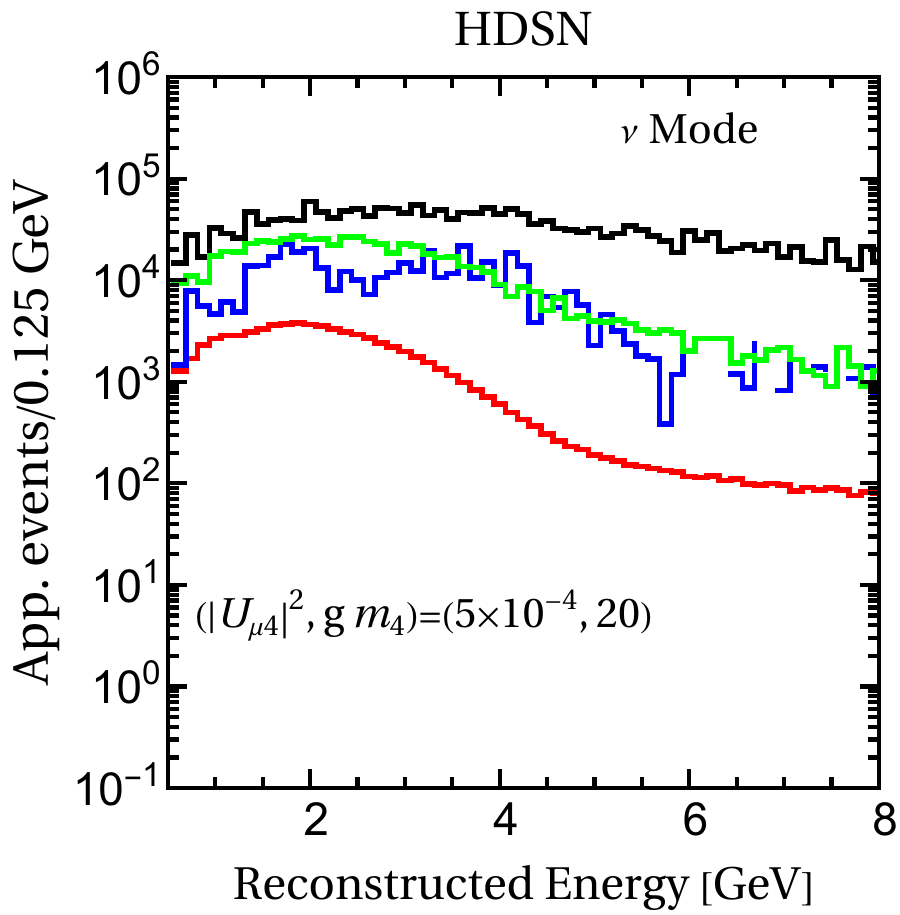}
\vskip .3cm
\includegraphics[height=4.7cm, width=4.7cm]{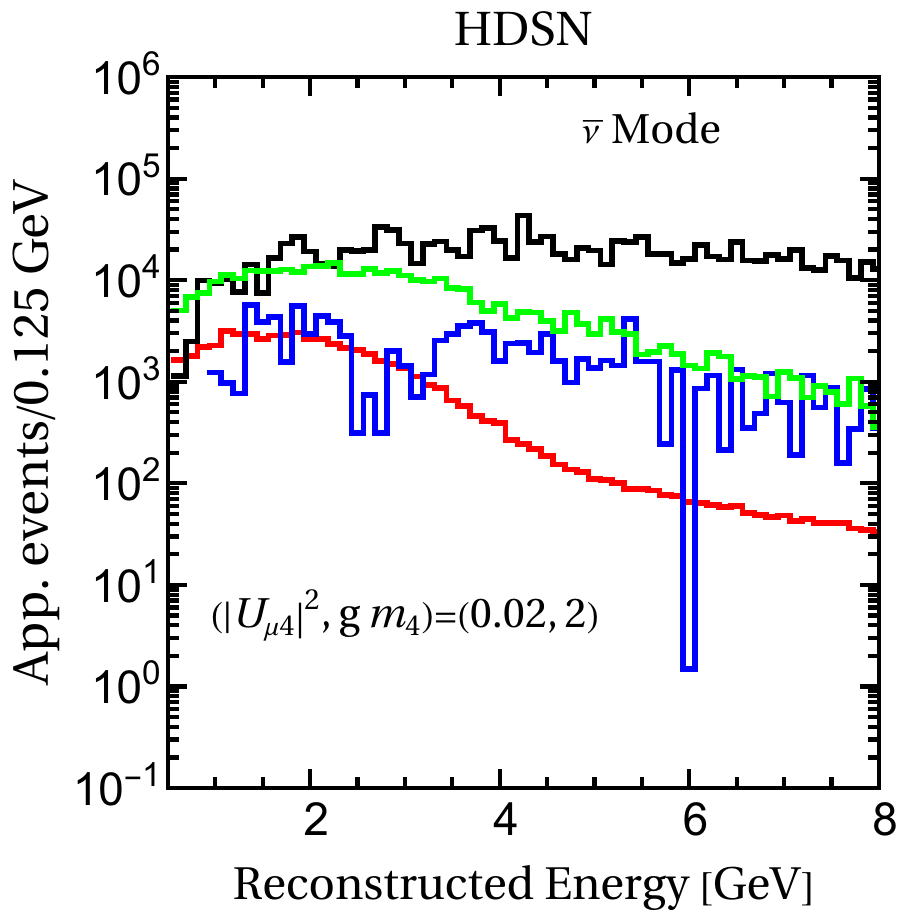}
\hskip .3cm
\includegraphics[height=4.7cm, width=4.7cm]{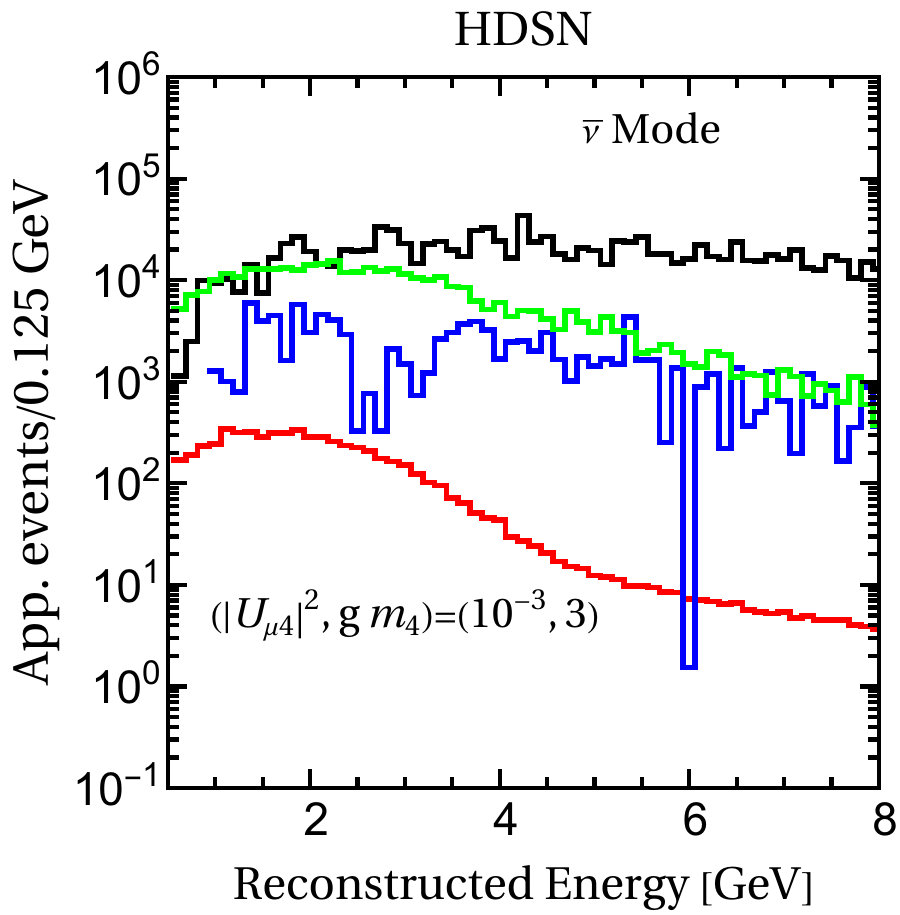}
\hskip .3cm
\includegraphics[height=4.7cm, width=4.7cm]{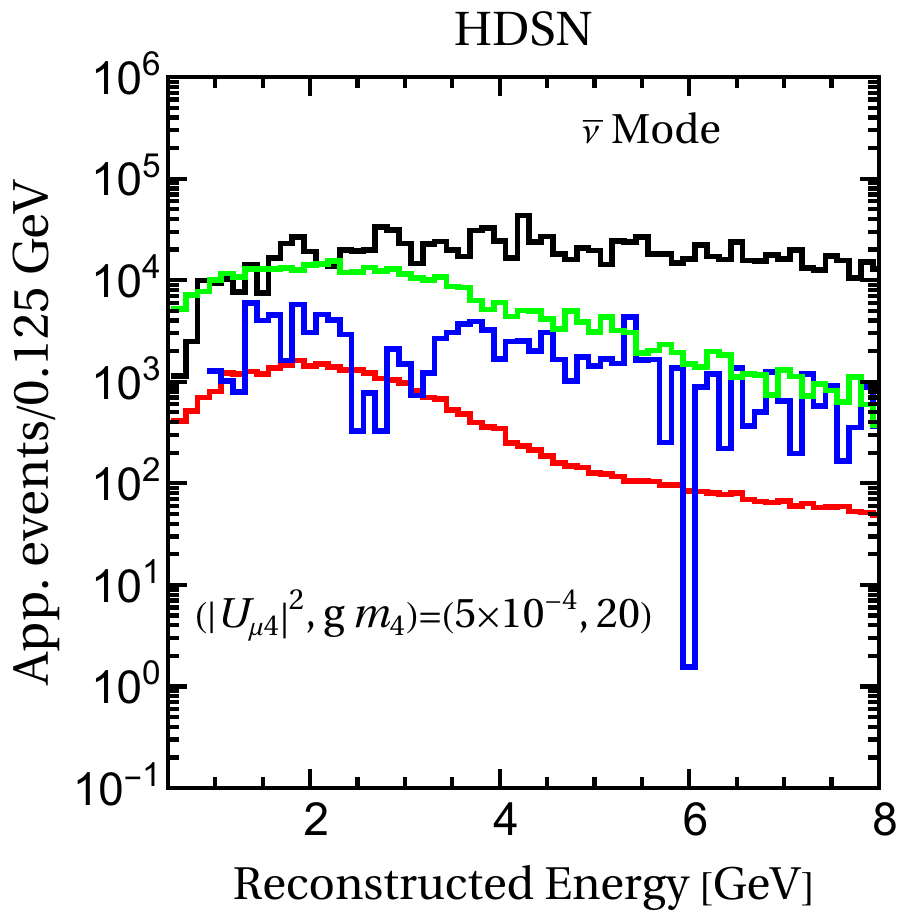}
\caption{Appearance event spectra expected at the DUNE ND-LAr detector for various choices of the
HDSN parameters $|U_{\mu 4}|^2$ and $g m_4$ (Dirac case), assuming a running time of 3.5 years on axis
in either neutrino or antineutrino mode.
The top and bottom panels correspond to the neutrino and antineutrino modes, respectively,
and the colour code is the same as in the left panels of Fig.~\ref{fig:ND_spectra_SM}.}
\label{fig:ND_appearance_HDSN}
\end{figure}
\begin{figure}
\centering
\includegraphics[height=6.7cm, width=6.7cm]{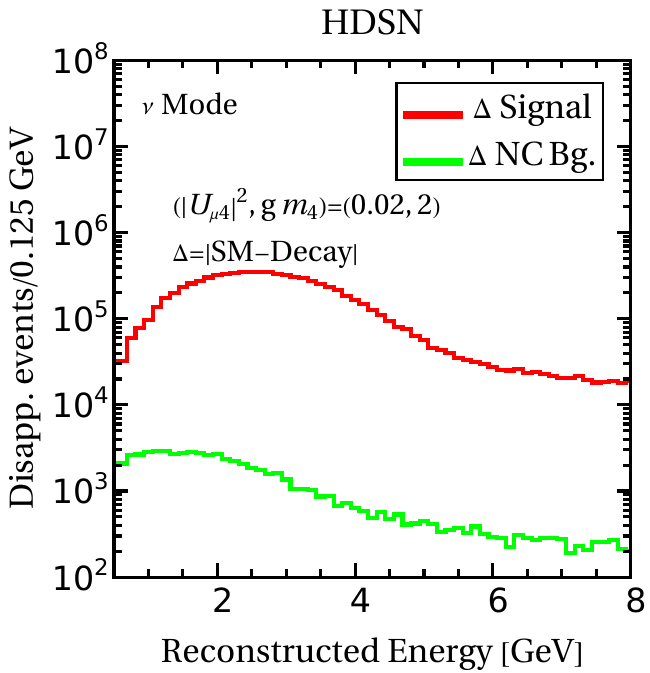}
\hskip 1cm
\includegraphics[height=6.7cm, width=6.7cm]{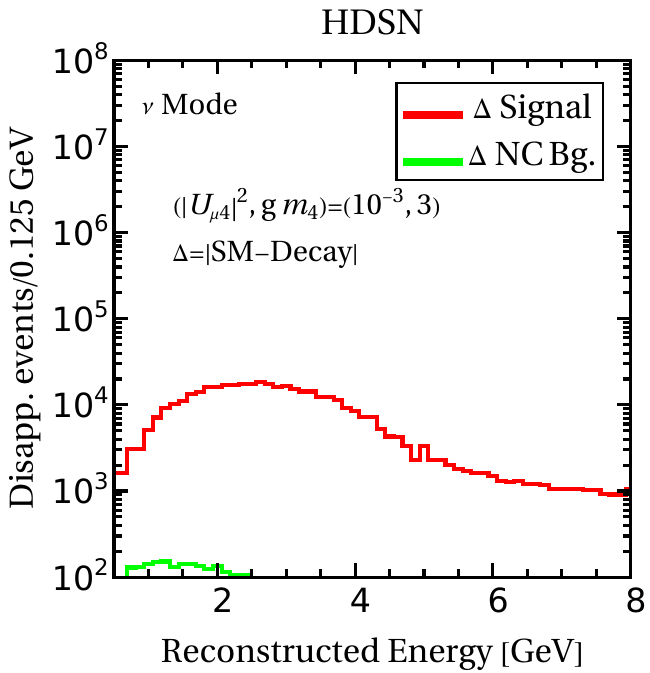}\\
\vskip .3cm
\includegraphics[height=6.7cm, width=6.7cm]{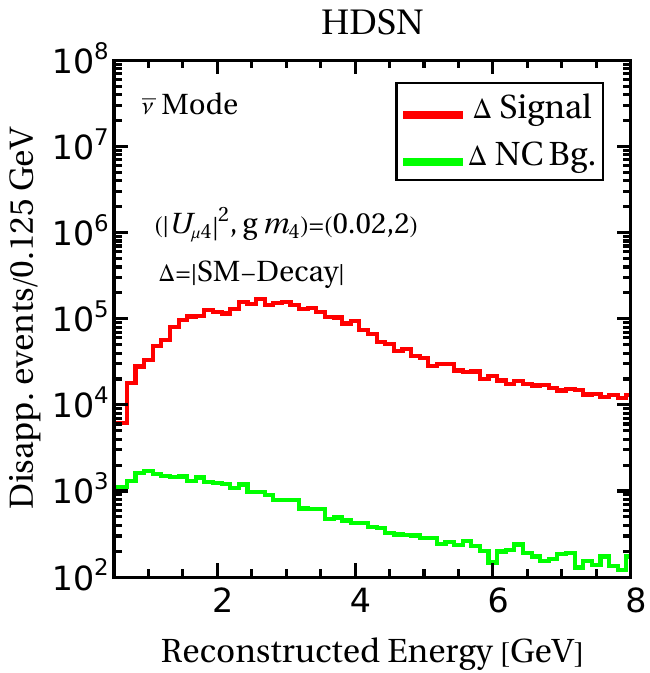}
\hskip 1cm
\includegraphics[height=6.7cm, width=6.7cm]{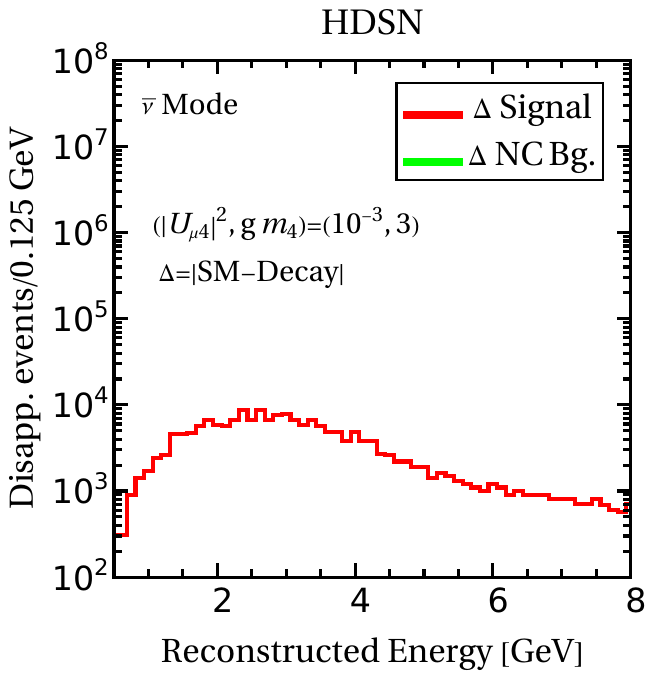}
\caption{Disappearance event spectra expected at the DUNE ND-LAr detector for various choices of the
HDSN parameters $|U_{\mu 4}|^2$ and $g m_4$ (Dirac case), assuming a running time of 3.5 years
on axis in either neutrino or antineutrino mode.
The red (green) curves show the difference between the number of signal (background) events
predicted in the SM and HDSN scenario as a function of the reconstructed (anti)neutrino energy.
The top and bottom panels correspond to the neutrino and antineutrino modes, respectively.}
\label{fig:ND_disappearance_HDSN}
\end{figure}

%%%%%%%%%%%%%%%%%%%%%%%%%%%%%%%%%%%%%%
\subsection{Appearance and disappearance spectra in the HDSN scenario}
%%%%%%%%%%%%%%%%%%%%%%%%%%%%%%%%%%%%%%
%
Let us now see how the signal and backgrounds are affected by the presence of a heavy decaying sterile neutrino.
Fig.~\ref{fig:ND_appearance_HDSN} displays the appearance spectra expected
at the DUNE ND-LAr detector for three different choices of the HDSN parameters,
assuming neutrinos are Dirac fermions. As in Fig.~\ref{fig:ND_spectra_SM}, a running time
of 3.5 years is assumed in both the neutrino (upper panels) and the antineutrino (lower panels) modes.
In the left panels, we took $(|U_{\mu 4}|^2, g m_4) = (0.02, 2\, \mbox{eV})$,
which can explain the LSND and MiniBooNE anomalies~\cite{deGouvea:2019qre} but has
been excluded at more than $3\sigma$ confidence level by MicroBooNE~\cite{Hostert:2024etd}.
In the middle and right panels, we chose the parameter values $(|U_{\mu 4}|^2, g m_4) = (10^{-3}, 3\, \mbox{eV})$
and $(5 \times 10^{-4}, 20\, \mbox{eV})$, respectively,
which can neither explain the LSND and MiniBooNE anomalies nor be tested by SBN~\cite{deGouvea:2019qre},
but, as we are going to see later, can be probed at the DUNE near detector.
In all three cases, the appearance signal is strongly enhanced with respect to the SM
by the decays of the heavy sterile neutrinos, while the misidentification and neutral current
backgrounds are only mildly affected\footnote{The intrinsic background is independent
of the physics scenario, and is therefore unchanged.}.
The number of appearance events depends on the HDSN parameters in a way that is well
explained by Formula~(\ref{eq:Pmue_Dirac}): $|U_{\mu 4}|^2$ controls the proportion of heavy sterile
neutrinos in the neutrino flux, while the value of $g m_4$ determines how many of these
decay to $\nu_e$'s
(the larger $g m_4$, the larger the fraction of sterile neutrinos that decay before reaching the detector,
since the sterile neutrino decay rate $\Gamma_4$ is proportional to $|g|^2 m^2_4$).
As a result, the appearance signal increases both with $|U_{\mu 4}|^2$ and $g m_4$.
As for backgrounds, the numbers of misidentification and neutral current events
in the HDSN scenario
deviate from the SM prediction by a fraction $2 |U_{\mu 4}|^2$ at most\footnote{This statement
also holds in the case of Majorana neutrinos, as can be checked from Eqs.~(\ref{eq:NC_bckgd_Majorana_nu})
and~(\ref{eq:NC_bckgd_Majorana_antinu}).},
as can be seen from Eqs.~(\ref{eq:misID_bckgd})\ and~(\ref{eq:NC_bckgd_Dirac}).
This explains why these backgrounds are only marginally affected by the presence of the heavy sterile neutrino.

Let us now consider the disappearance channel.
The left and right panels of Fig.~\ref{fig:ND_disappearance_HDSN} show the difference
between the predictions of the SM and HDSN scenario for the numbers of signal events (red curves)
and neutral current background events (green curves) for two different choices of model parameters,
assuming neutrinos are Dirac fermions.
As in the previous figures, the upper and lower panels correspond to the neutrino and antineutrino modes,
respectively, and a running time of 3.5 years is assumed for each mode.
In the left panels, we chose the same parameter values as in the left panels of
Fig.~\ref{fig:ND_appearance_HDSN}, namely $(|U_{\mu 4}|^2, g m_4) = (0.02, 2\, \mbox{eV})$,
while we took $(|U_{\mu 4}|^2, g m_4) = (10^{-3}, 3\, \mbox{eV})$ in the right panels,
as in the middle panels of Fig.~\ref{fig:ND_appearance_HDSN}.
At variance with the appearance signal, the disappearance signal
is only mildly affected by the presence of the heavy sterile neutrino (unless one considers
values of $|U_{\mu 4}|^2$ larger than the MINOS 90\% C.L. upper bound).
This can easily be understood from the survival probability~(\ref{eq:Pmumu}), which deviates
from the SM value $P^{\rm SM}_{\mu \mu} = P^{\rm SM}_{\bar \mu \bar \mu} \simeq 1$
by a fraction $2 |U_{\mu 4}|^2$ at most.
The same holds for the neutral current background,
which according to Eq.~(\ref{eq:NC_bckgd_Dirac}) is reduced relative to the SM by a factor
ranging from\footnote{In the case of Majorana neutrinos, this factor ranges from $\approx 1 - 2 |U_{\mu 4}|^2$ 
to $\approx 1 - \frac 5 4 |U_{\mu 4}|^2$ in the neutrino mode (resp. $\approx 1 - \frac 1 2 |U_{\mu 4}|^2$
in the antineutrino mode),
as can be checked from Eqs.~(\ref{eq:NC_bckgd_Majorana_nu}) and~(\ref{eq:NC_bckgd_Majorana_antinu}).}
$\simeq 1 - 2 |U_{\mu 4}|^2$ to $\simeq 1 - |U_{\mu 4}|^2$, depending on the value of $g m_4$.

%%%%%%%%%%%%%%%%%%%%%%%%%%%%%%
\subsection{Sensitivity of ND-LAr to the HDSN parameters}
%%%%%%%%%%%%%%%%%%%%%%%%%%%%%%

\begin{figure}
\centering
\includegraphics[height=7cm, width=7cm]{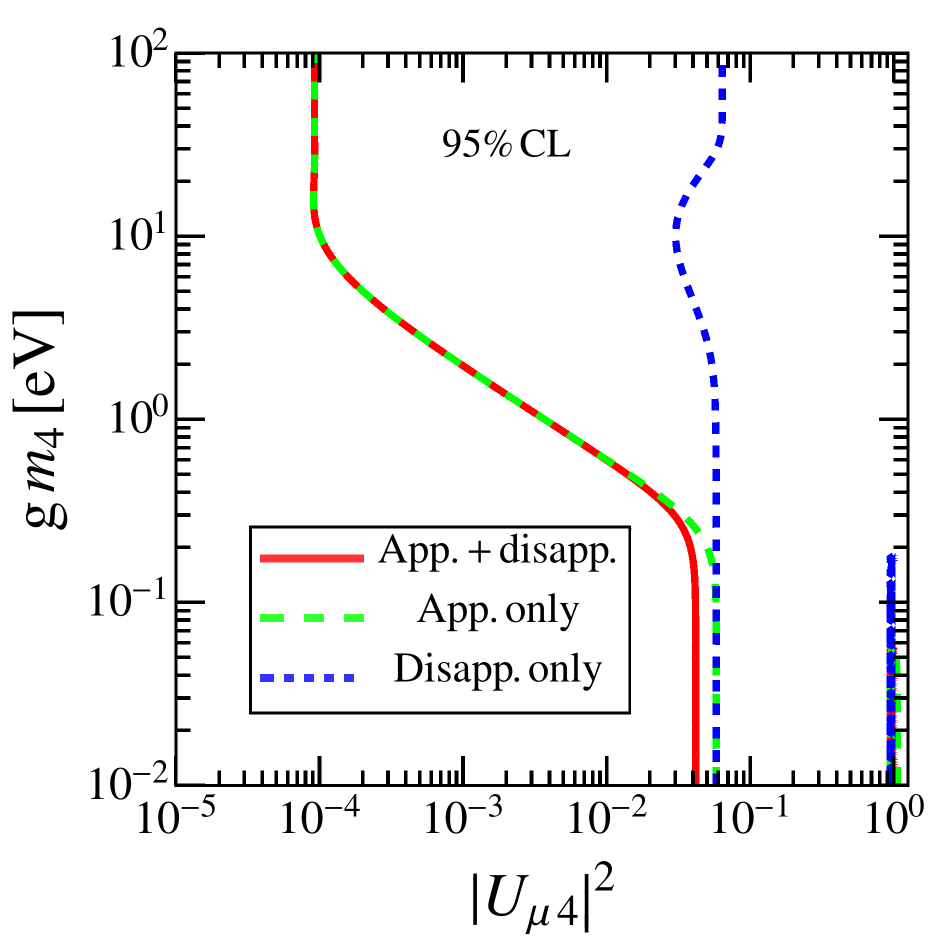}
\includegraphics[height=7cm, width=7cm]{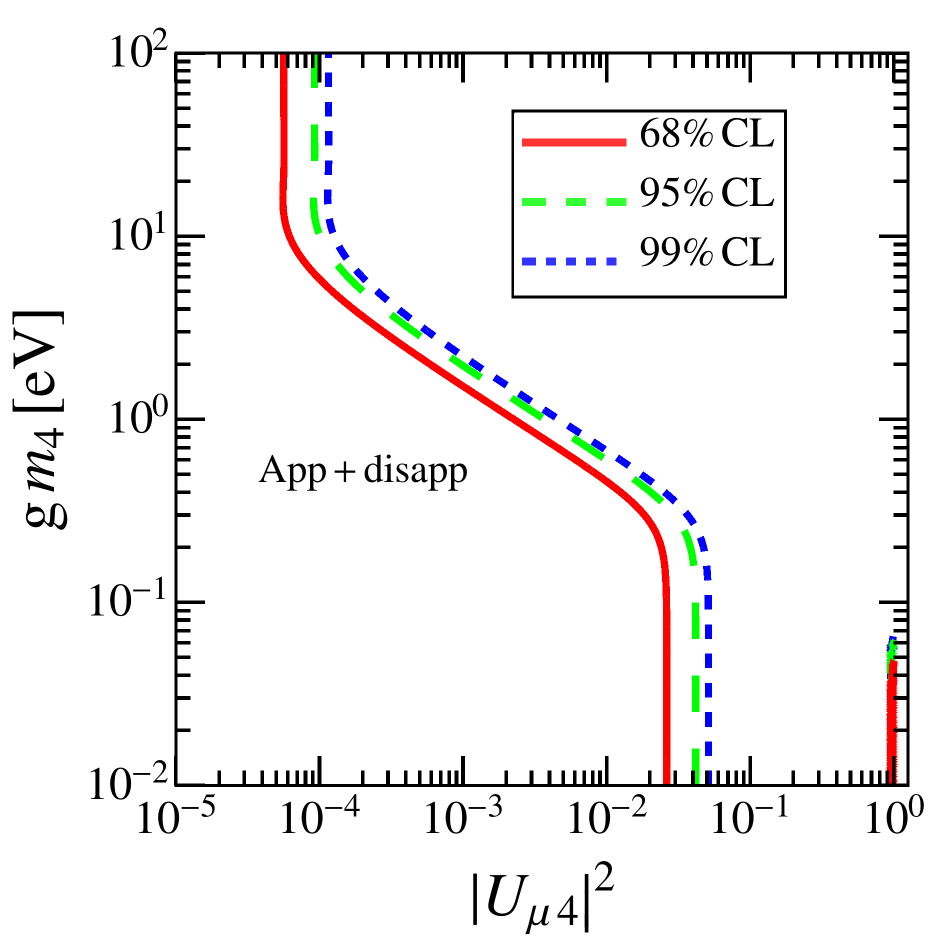}
\caption{Expected ND-LAr sensitivity to the heavy decaying sterile neutrino parameters $|U_{\mu 4}|^2$ and $g m_4$ (Dirac case),
assuming 3.5 years of on-axis data taking in both neutrino and antineutrino modes, and 10\% signal and background uncertainties.
The left plot shows the 95\% C.L. sensitivities obtained using appearance data only (green long-dashed curve),
disappearance data only (blue short-dashed curve), and both appearance and disappearance data (red solid curve).
The right plot compares the appearance + disappearance combined sensitivities at various confidence levels.
The regions that are expected to be excluded by future data are enclosed within these curves.}
\label{fig:sensitivity_on_axis}
\end{figure}

We now move on to assess the ability of the DUNE ND-LAr detector to test the HDSN scenario,
focusing first on the case of Dirac neutrinos.
The left plot of Fig.~\ref{fig:sensitivity_on_axis} displays the expected 95\% C.L. sensitivity regions
in the $(|U_{\mu 4}|^2, g m_4)$ plane corresponding to appearance data only (green long-dashed curve),
disappearance data only (blue short-dashed curve) and to the combination of appearance and
disppearance data (red solid curve). We assumed 3.5 years of on-axis data taking in both the neutrino
and antineutrino modes, as well as 10\% normalization uncertainties for both signal
and background. Comparing the two dashed curves shows that the ND-LAr
sensitivity to the mixing parameter $U_{\mu 4}$ is dominated by appearance data, except
for values of $g m_4$ smaller than a few $0.1\, \mbox{eV}$, where the appearance and disappearance
channels are equally efficient in constraining $|U_{\mu 4}|^2$. Combining both channels in this region
improves the expected upper bound on $|U_{\mu 4}|^2$, without making it
competitive with the MINOS constraint ($|U_{\mu 4}|^2 < 0.023$ at 90\% C.L.).
Finally, the right plot of Fig.~\ref{fig:sensitivity_on_axis} compares the appearance + disappearance
combined sensitivities at different confidence levels (68\%, 95\% and 99\% C.L.).

The different shapes of the appearance and disappearance sensitivity curves in the left plot
of Fig.~\ref{fig:sensitivity_on_axis} can be understood from the formulae of Section~\ref{sec:NuDecay}.
For the disappearance channel, the difference between the predictions of the SM and HDSN
scenario for the signal and background scales as $|U_{\mu 4}|^2$, as already noticed,
with little dependence on $g m_4$ (see Eqs.~(\ref{eq:Pmumu}) and~(\ref{eq:NC_bckgd_Dirac}),
where the $g m_4$ dependence is hidden in the sterile neutrino decay rate $\Gamma_4$).
In fact, the sensitivity to $|U_{\mu 4}|^2$ in the disappearance channel is almost
independent of $g m_4$, except around $g m_4 \sim 10\, \mbox{eV}$. The increased sensitivity
to $|U_{\mu 4}|^2$ in this region arises
from the energy dependence of the disappearance signal, which is due to the subleading term 
$|U_{\mu 4}|^4\, e^{- \Gamma_4 L}$ in $P_{\mu \mu}$ and is maximal for $g m_4 \sim 10\, \mbox{eV}$.
For the appearance channel, instead, the signal strongly depends on $g m_4$.
For large values of this parameter, corresponding to a large $\Gamma_4$, most
sterile neutrinos decay before reaching the DUNE near detector and the appearance
probability~(\ref{eq:Pmue_Dirac}) reduces to $P_{\mu e} \simeq |U_{\mu 4}|^2$. This explains
the upper part of the green long-dashed curve. For smaller values of $g m_4$, hence of $\Gamma_4$,
only a fraction of the sterile neutrinos decay, leading to a suppression of the appearance
probability by a factor $(1 -  e^{-\Gamma_4 L})$.
As a result, the sensitivity to $|U_{\mu 4}|^2$ decreases with $g m_4$.
Finally, for $g m_4$ small enough, most sterile neutrinos reach the near detector before decaying,
resulting in a strong suppression of appearance events. The sensitivity to $|U_{\mu 4}|^2$
is then driven by neutral current and misidentification background events,
whose number is suppressed relative to the SM by a factor $\simeq 1 - 2 |U_{\mu 4}|^2 (1 - |U_{\mu 4}|^2)$
in the low $g m_4$ region (as can be seen by taking the limit $e^{- \Gamma_4 L} \to 1$ in Eqs.~(\ref{eq:misID_bckgd})
and~(\ref{eq:NC_bckgd_Dirac})). 
This explains the lower part of the exclusion curve, as well as
the narrow allowed region\footnote{This narrow region is also allowed by disappearance data,
since both the signal, Eq.~(\ref{eq:Pmumu}), and the background, Eq.~(\ref{eq:NC_bckgd_Dirac}),
are suppressed by a factor $\simeq 1 - 2 |U_{\mu 4}|^2 (1 - |U_{\mu 4}|^2)$ relative to the SM
in the limit $e^{- \Gamma_4 L} \to 1$.} close to $|U_{\mu 4}|^2 = 1$.
However, as already mentioned, the sensitivity of the DUNE ND-LAr detector in the low $g m_4$
region is not competitive with the MINOS upper bound.

%%%%%%%%%%%%%%%%%%%%%%%
\subsection{Dirac versus Majorana neutrinos}
%%%%%%%%%%%%%%%%%%%%%%%

So far we assumed that neutrinos were Dirac fermions.
The signatures of a heavy decaying sterile Majorana neutrino in the DUNE near detector
are qualitatively similar, but not identical, as we discuss below.
The differences are mainly due to the fact that sterile Majorana neutrinos have lepton number violating
decay modes, resulting in the additional
appearance channels $\nu_\mu \to \bar \nu_e$ and $\bar \nu_\mu \to \nu_e$,
with the same probabilities as the lepton number conserving transitions
$\nu_\mu \to \nu_e$ and $\bar \nu_\mu \to \bar \nu_e$ (the only allowed ones in the Dirac case).
Incidentally, this implies that 
the sterile neutrino decay rate is twice as large in the Majorana case
as it is in the Dirac case (i.e., $\Gamma^M_4 = 2 \Gamma^D_4$).
\begin{figure}
\centering
\includegraphics[height=10cm, width=10cm]{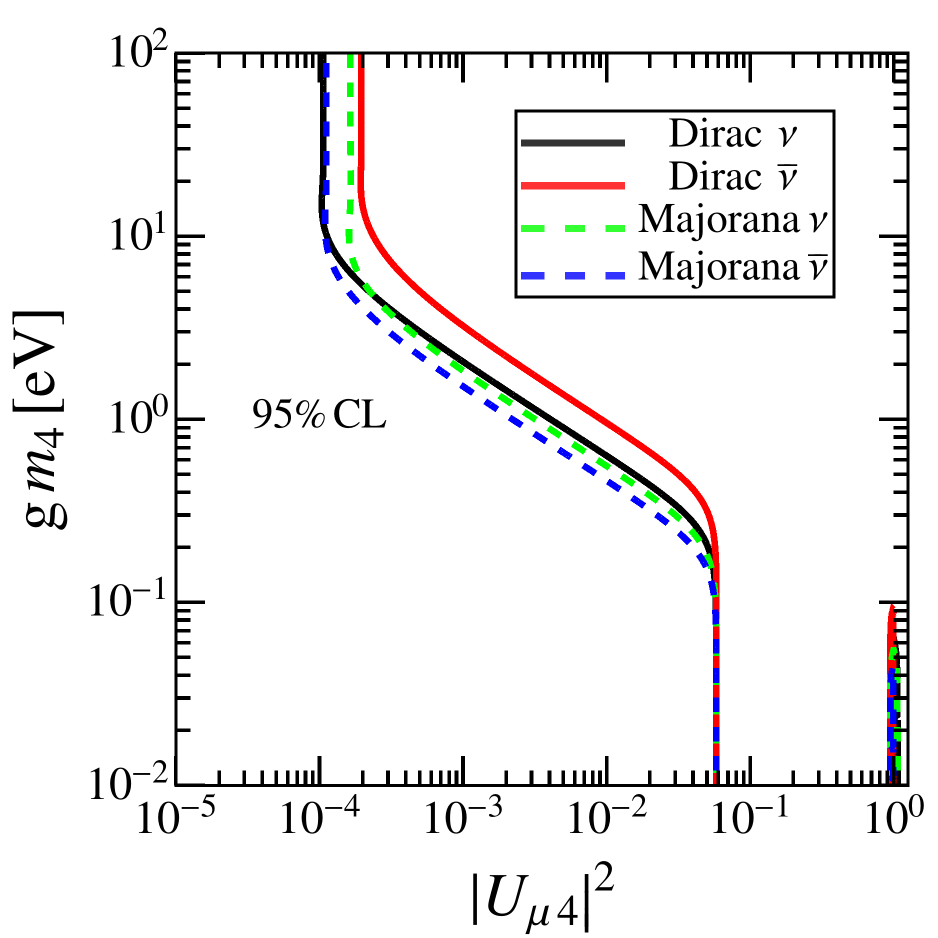}
\caption{Impact of the neutrino nature and of the running mode (neutrino vs. antineutrino flux) on the expected
95\% C.L. (combined appearance and disappearance) sensitivity of ND-LAr to the heavy decaying
sterile neutrino parameters $|U_{\mu 4}|^2$ and $g m_4$.
Black solid curve: Dirac case, neutrino mode. Red solid curve: Dirac case, antineutrino mode.
Green dashed curve: Majorana case, neutrino mode. Blue dashed curve: Majorana case, antineutrino mode.
Each sensitivity curve assumes 3.5 years of on-axis data taking and 10\% signal and background uncertainties.}
\label{fig:sensitivity_neutrino_nature}
\end{figure}

Fig.~\ref{fig:sensitivity_neutrino_nature} shows how the expected 95\% C.L. sensitivity
of ND-LAr to the HDSN parameters depends on the neutrino nature and on the running mode
(neutrino vs. antineutrino flux).
Each sensitivity curve corresponds to one of the four possibilities:
Dirac case, neutrino mode (black solid curve); Dirac case, antineutrino mode (red solid curve);
Majorana case, neutrino mode (green dashed curve); Majorana case, antineutrino mode (blue dashed curve).
The sensitivities were computed using both appearance and disappearance data,
assuming 3.5 years of on-axis data taking and 10\% normalization uncertainties for signal and background.
One can clearly see some differences between the four curves at large and intermediate
values of $g m_4$. To understand them, it is useful to consider how the
signal event rates are affected by the neutrino nature and by the running mode.
Let us first consider the large $g m_4$ region ($g m_4 > 10\, {\rm eV}$), where the sensitivity
is dominated by appearance data, and almost all sterile (anti)neutrinos decay before reaching the detector.
Given that {\it (i)} sterile (anti)neutrinos only decay to electron (anti)neutrinos in the Dirac case;
{\it (ii)} both sterile neutrinos and antineutrinos decay
to $\nu_e$ and $\bar \nu_e$ with equal probabilities in the Majorana case;
and {\it (iii)} the charged current cross section is larger for neutrinos than for antineutrinos,
one has
\begin{equation}
  S^D_{\bar \nu} < S^M_{\bar \nu} \simeq S^M_\nu < S^D_\nu\, ,
\label{eq: S_B_large_gm4}
\end{equation}
where $S^D_\nu$ ($S^M_\nu$) represents the total appearance signal rate in the neutrino
mode, assuming neutrinos are Dirac (Majorana) fermions, and analogously
for $S^D_{\bar \nu}$ ($S^M_{\bar \nu}$) in the antineutrino mode.
The small difference between $S^M_\nu$ and $S^M_{\bar \nu}$ is due to the fact that
the total $\nu_\mu +\bar \nu_\mu$ flux is smaller in the antineutrino mode than in the neutrino mode.
Eq.~(\ref{eq: S_B_large_gm4}), together with the fact that the background level is higher
in the neutrino mode than in the antineutrino mode, explains the differences in the sensitivity
curves that can be observed in the large $g m_4$ region of Fig.~\ref{fig:sensitivity_neutrino_nature}.
In particular, in the Dirac case, the neutrino mode can probe smaller values of $|U_{\mu 4}|^2$
than the antineutrino mode.
As for the Majorana case, although the signal is roughly the same in the neutrino and antineutrino modes,
the antineutrino mode has a significantly better sensitivity to $|U_{\mu 4}|^2$ due to the lower background rate.
For the same reason, the sensitivity to $|U_{\mu 4}|^2$ is practically the same for the Dirac case
in the neutrino mode as for the Majorana case in the antineutrino mode, while it is slightly better
for the Majorana case in the neutrino mode than for the Dirac case in the antineutrino mode.

In the intermediate $g m_4$ region ($0.2\, {\rm eV} \lesssim g m_4 \lesssim 10\, {\rm eV}$),
only a fraction $1 - e^{- \Gamma_4 L}$ of the sterile (anti)neutrinos decays before reaching
the DUNE near detector. Since $\Gamma^M_4 = 2 \Gamma^D_4$, there are twice as many
decays in the Majorana case as in the Dirac case, resulting in the following hierarchy of signal rates:
\begin{equation}
  S^D_{\bar \nu} < S^D_\nu < S^M_\nu \simeq S^M_{\bar \nu} \, .
\label{eq: S_B_intermediate_gm4}
\end{equation}
One recovers the fact that, in the Dirac case, the sensitivity to $|U_{\mu 4}|^2$ is higher
in the neutrino mode than in the antineutrino mode, but now smaller values of the mixing angle
can be probed in the Majorana case, especially in the antineutrino mode in which
the background level is lower (see Fig.~\ref{fig:sensitivity_neutrino_nature}).
Finally, in the low $g m_4$ region ($g m_4 \lesssim 0.1\, {\rm eV}$), 
almost all sterile (anti)neutrinos reach
the detector before decaying, and the appearance signal can be neglected. The sensitivity
to $|U_{\mu 4}|^2$ is then driven by the disappearance signal, and by the neutral current
and misidentification backgrounds in the appearance channel, all of which become
independent of the neutrino nature and of the running mode 
in the low $g m_4$ region\footnote{Indeed,
$P_{\mu \mu} = P_{\bar \mu \bar \mu}$ and, in the limit where sterile neutrino decays
can be neglected (which corresponds to \mbox{$e^{- \Gamma_4 L} \to 1$}),
$P_{\mu \mu}$ becomes independent of $\Gamma_4$, hence of the neutrino nature.
The same holds for the ratio $N^{\rm HDSN}_{\rm mis-ID} / N^{\rm SM}_{\rm mis-ID}$,
given by $P_{\mu \mu}$, and for $N^{\rm HDSN}_{\rm NC bckgd} / N^{\rm SM}_{\rm NC bckgd}$,
which reduces to the same expression as $P_{\mu \mu}$ when $e^{- \Gamma_4 L} \to 1$.}.
As a result, the sensitivity to $|U_{\mu 4}|^2$ in the low $g m_4$ region does not depend on the neutrino nature
nor on the running mode (up to the differences between the neutrino and antineutrino fluxes), as can be seen in Fig.~\ref{fig:sensitivity_neutrino_nature}.

The main conclusion one can draw from Fig.~\ref{fig:sensitivity_neutrino_nature} is that it is possible,
at least in principle, to distinguish between Dirac and Majorana neutrinos if the HDSN scenario
is realized in Nature with $g m_4 \gtrsim {\rm few}\, 0.1\, {\rm eV}$. The key observables for this purpose
are the numbers of appearance events (or more precisely, the appearance spectra)
in the neutrino and antineutrino modes.
If neutrinos are Dirac fermions, one expects ND-LAr to observe a larger excess of appearance events
in the neutrino mode than in the antineutrino mode, while the opposite is true in the Majorana case.
However, telling Dirac from Majorana neutrinos might be challenging in terms of statistics
(especially if $|U_{\mu 4}|^2$ is small), and more than 3 years of data taking at ND-LAr
would probably be required.
A more promising avenue may be to consider a magnetized detector (such as SAND,
one of the three near detectors envisaged by the DUNE collaboration) and to reconstruct separately
the electron and positron spectra in the neutrino and antineutrino modes. In the Dirac case,
only $\nu_e$ signal events should be present in the neutrino mode
(and $\bar \nu_e$ signal events in the antineutrino mode), while both $\nu_e$ and $\bar \nu_e$ signal events
are expected in the Majorana case, in approximately
the same proportions in the neutrino and antineutrino modes.
We plan a detailed study of the possibility to distinguish between Dirac
and Majorana neutrinos in the HDSN scenario in a future work.

%%%%%%%%%%%%%%%%%%%%%%%%%%%%%%%%%%%%%%%%%%%%%%%%%
\subsection{Comparing the ND-LAr and MicroBooNE/SBN sensitivities to the HDSN parameters}
\begin{figure}
\centering
\includegraphics[height=7.4cm, width=7.4cm]{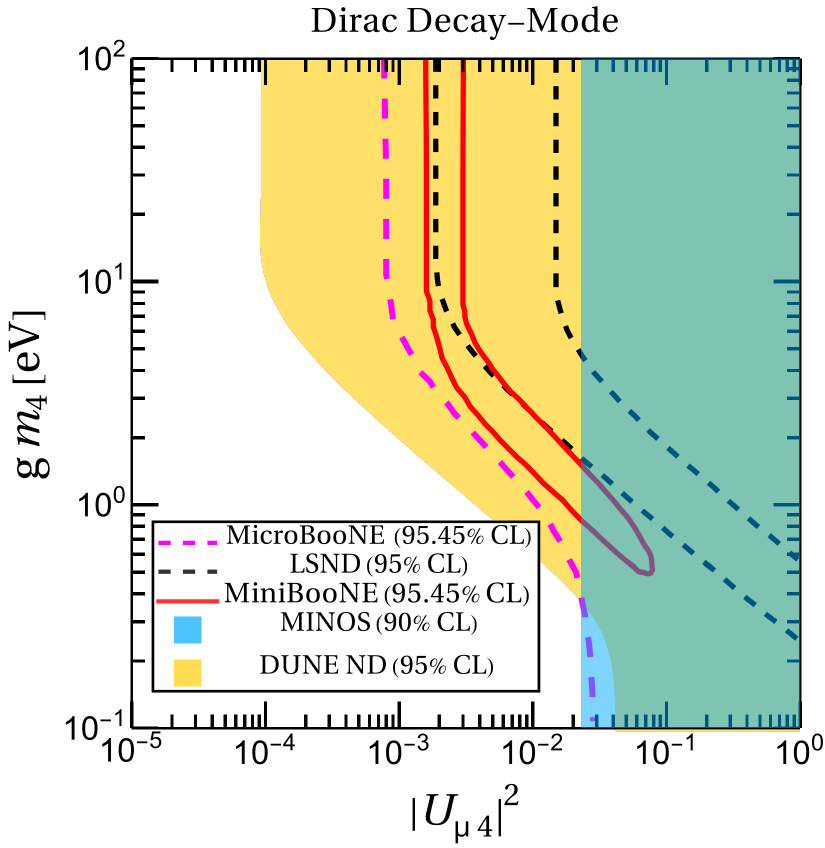}
\includegraphics[height=7.4cm, width=7.4cm]{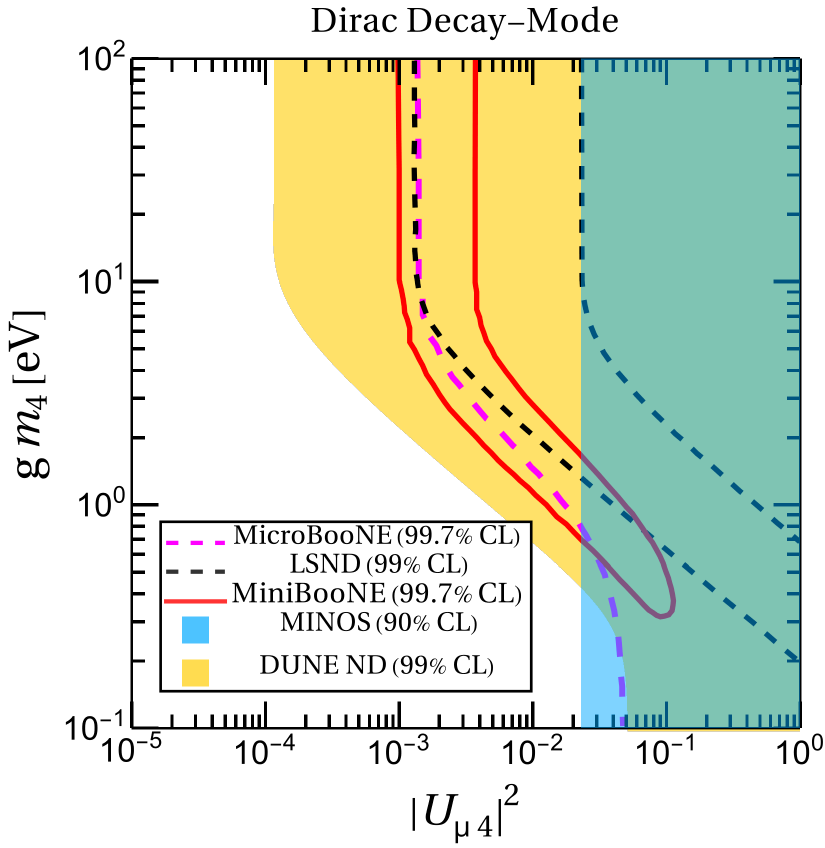}
\caption{Comparison of the 95\% C.L. (left) and 99\% C.L. (right) expected sensitivity of ND-LAr to the HDSN
parameters $|U_{\mu 4}|^2$ and $g m_4$ (Dirac case) with the regions consistent with the LSND~\cite{deGouvea:2019qre}
and MiniBooNE~\cite{Hostert:2024etd} anomalies, and with the region excluded by the MicroBooNE
experiment~\cite{Hostert:2024etd}. For ND-LAr, we assumed 3.5 years of on-axis data taking in both neutrino
and antineutrino modes, as well as 10\% signal and background uncertainties.
The confidence levels for the MiniBooNE and MicroBooNE regions, taken from Ref.~\cite{Hostert:2024etd},
are $2\sigma$ in the left plot and $3\sigma$ in the right plot.
Also shown is the 90\% C.L. MINOS/MINOS+ upper bound on $|U_{\mu 4}|^2$~\cite{MINOS:2017cae}.}
\label{fig:sensitivity_LSND_MB_Dirac}
\end{figure}

\begin{figure}
\centering
\includegraphics[height=7.4cm, width=7.4cm]{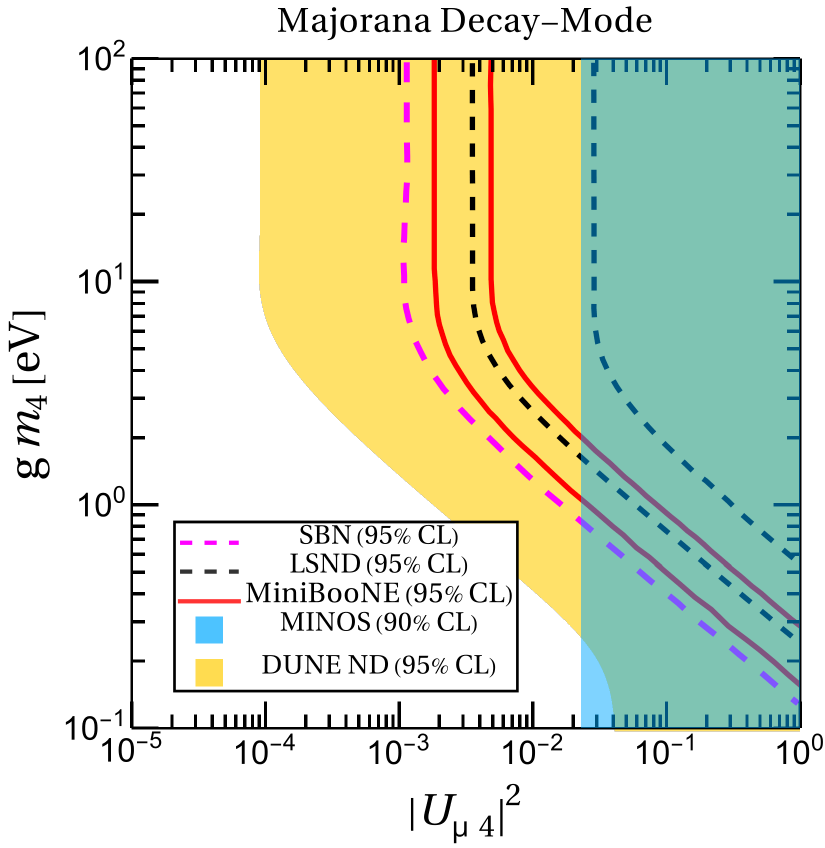}
\includegraphics[height=7.4cm, width=7.4cm]{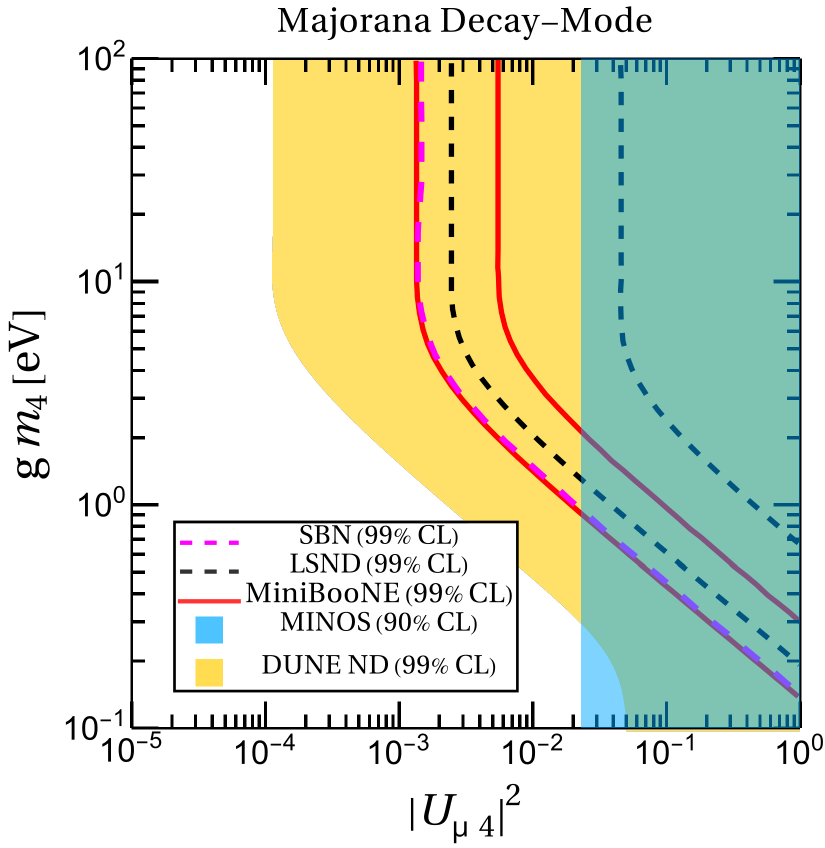}
\caption{Comparison of the 95\% C.L. (left) and 99\% C.L. (right) expected sensitivity of ND-LAr to the HDSN
parameters $|U_{\mu 4}|^2$ and $g m_4$ (Majorana case) with the regions consistent with the LSND
and MiniBooNE anomalies, and with the expected sensitivity of the SBN experiment.
For ND-LAr, we assumed 3.5 years of on-axis data taking in both neutrino
and antineutrino modes, as well as 10\% signal and background uncertainties.
The LSND, MiniBooNE and SBN regions are taken from Ref.~\cite{deGouvea:2019qre}.
Also shown is the 90\% C.L. MINOS/MINOS+ upper bound on $|U_{\mu 4}|^2$~\cite{MINOS:2017cae}.}
\label{fig:sensitivity_LSND_MB_Majorana}
\end{figure}

%%%%%%%%%%%%%%%%%%%%%%%%%%%%%%%%%%%%%%%%%%%%%%%%%
%
Finally, in Figs.~\ref{fig:sensitivity_LSND_MB_Dirac} and~\ref{fig:sensitivity_LSND_MB_Majorana},
we compare the expected ND-LAr sensitivity to the HDSN parameters
with the regions consistent with the LSND and MiniBooNE anomalies,
and with the current constraint from the MicroBooNE experiment (Dirac case) or the expected sensitivity
of the SBN program at Fermilab (Majorana case).
For ND-LAr, we considered 3.5 years of on-axis data taking in both neutrino and antineutrino modes.
Appearance and disappearance data were combined, and 10\% signal and background uncertainties were assumed.
Fig.~\ref{fig:sensitivity_LSND_MB_Dirac} corresponds to the Dirac case, with the 95\% C.L.
and 99\% C.L. expected ND-LAr sensitivities shown in the left and right panels, respectively.
In Ref.~\cite{Hostert:2024etd}, an updated fit of MiniBooNE neutrino and antineutrino data
within the HDSN scenario was performed, and constraints on the HDSN parameters were derived
using recent data from the MicroBooNE experiment. The corresponding $2\sigma$ and $3\sigma$
regions are shown on the left and right plots, respectively, where the area allowed by MiniBooNE
is within the red solid contour, and the region excluded by MicroBooNE lies to the right of
the pink long-dashed curve. Also shown is the LSND allowed
region, taken from Ref.~\cite{deGouvea:2019qre}.
While the current MicroBooNE data already excludes most of the MiniBooNE
allowed region at the $3\sigma$ level, ND-LAr will be able to probe a much larger region of the HDSN
parameter space, and to fully exclude this scenario as a solution of the LSND and
MiniBooNE anomalies. In addition, its sensitivity might prove crucial to confirm or reject a possible hint
of $\nu_e$ appearance in future MicroBooNE data.
Fig.~\ref{fig:sensitivity_LSND_MB_Majorana} corresponds to the Majorana case, which was
not considered\footnote{The authors of Ref.~\cite{Hostert:2024etd} did not consider
the Majorana case in order to avoid conflict with the stringent experimental limits on the flux of $\bar \nu_e$
from the Sun~\cite{Hostert:2020oui}. However, the constraints discussed in Ref.~\cite{Hostert:2020oui}
actually apply to the HDSN scenario of Ref.~\cite{Dentler:2019dhz}, in which the electron
neutrino contains a $\nu_4$ component and the light scalar field $\phi$ can decay to
a neutrino-antineutrino pair, leading to $\nu_4 \to \nu_i \phi \to \nu_i \nu_j \bar \nu_k$ ($i,j,k = 1,2,3$) in the Sun.
In the scenario of Ref.~\cite{deGouvea:2019qre} considered in this paper,
$\phi$ is extremely light and does not decay to active neutrinos.
While decays $\nu_4 \to \bar \nu_e \phi$ are possible in the Majorana case, solar neutrinos
are produced in a combination of matter Hamiltonian eigenstates that only very weakly mix with $\nu_4$,
thus evading the constraints on the $\bar \nu_e$ flux from the Sun.}
in Ref.~\cite{Hostert:2024etd}. 
We therefore show the LSND and MiniBooNE
allowed regions from Ref.~\cite{deGouvea:2019qre}, as well as the expected sensitivity
of the SBN experiment (of which MicroBooNE is one of the three detectors) from the same paper.
The same conclusions as in the Dirac case hold.
Namely, ND-LAr will be able to probe a much larger region of the HDSN
parameter space than SBN, and either to fully exclude this scenario as a solution of the LSND
and MiniBooNE anomalies, or to confirm a possible hint of a positive signal
in future MicroBooNE, SBND or ICARUS data.
Notice that the shape of the MiniBooNE allowed region given
by Ref.~\cite{deGouvea:2019qre} and shown in Fig.~\ref{fig:sensitivity_LSND_MB_Majorana}
is significantly different from the one of Ref.~\cite{Hostert:2024etd}, displayed in
Fig.~\ref{fig:sensitivity_LSND_MB_Dirac}.
This is due to the fact that only appearance data was considered
in Ref.~\cite{deGouvea:2019qre}, while both appearance and disappearance data were included
in the fit of Ref.~\cite{Hostert:2024etd}. In practice, however, the main difference between
the MiniBooNE allowed regions of Figs.~\ref{fig:sensitivity_LSND_MB_Dirac} and~\ref{fig:sensitivity_LSND_MB_Majorana}
lies in the part of the parameter space that is already excluded by MINOS/MINOS+.
The same comment applies to the sensitivity region of the SBN experiment
shown in Fig.~\ref{fig:sensitivity_LSND_MB_Majorana},
which does not take into account disappearance data, at variance with the region excluded by MicroBooNE
in Fig.~\ref{fig:sensitivity_LSND_MB_Dirac}.
One can also notice that, in the large $g m_4$ region of the parameter space, where the sensitivity
to $|U_{\mu 4}|^2$ is driven by appearance data, the MicroBooNE constraint is already as good
as the expected sensitivity of the SBN program
reported in Ref.~\cite{deGouvea:2019qre}.
This suggests that the final sensitivity of SBN will be better than indicated by the pink long-dashed
curve of Fig.~\ref{fig:sensitivity_LSND_MB_Majorana}.

Throughout this paper, we assumed that ND-LAr was located on the beam axis,
although it will be a movable detector, able to collect data at different angles from the on-axis position.
As discussed in Section~\ref{sec:Detector} and illustrated in Fig.~\ref{fig:nuSpec},
the value of the off-axis angle strongly affects the neutrino flux at the near detector.
Namely, the intensity of the flux decreases as the off-axis angle increases, and its peak
is shifted towards lower energies.
We have checked that this flux reduction weakens the sensitivity of ND-LAr to $|U_{\mu 4}|^2$,
as could have been anticipated.
This is the reason why we did not present any results for off-axis locations of the DUNE ND-LAr detector.
On the other hand, future off-axis data may help to constrain the energy dependence of the flux
and cross section uncertainties,
and make it possible to check how our results are affected by this energy dependence.

%%%%%%%%%%%%%%%%%%%%%%%%
\section{Conclusions}
\label{sec:conclusions}
%%%%%%%%%%%%%%%%%%%%%%%%

In this paper, we investigated the possibility to test the heavy decaying sterile neutrino hypothesis
at the DUNE liquid argon near detector (ND-LAr). More specifically, we considered the scenario
of Refs.~\cite{Palomares-Ruiz:2005zbh,deGouvea:2019qre}, in which a fourth, mostly sterile neutrino with a mass
in the keV-MeV range and a small muon neutrino component decays to an electron neutrino and an invisible
light scalar field, thus mimicking the excesses observed by the LSND and MiniBooNE experiments.
We showed that ND-LAr can probe a larger region of the heavy decaying sterile neutrino parameter space 
than the Fermilab SBN program,
and will be able to fully exclude this scenario as a solution to the LSND and MiniBooNE anomalies.
For instance, ND-LAr can exclude $|U_{\mu4}|^2 \gtrsim 10^{-4}$ for $g m_4 \geq 10\, \mbox{eV}$
at more than $95\%$ C.L. with 3.5 years of on-axis data taking in both neutrino and antineutrino modes.
This sensitivity might prove crucial to confirm or reject a possible hint
of $\nu_e$ appearance in future MicroBooNE, SBND or ICARUS data.

We also showed that it may be possible, in case of a positive signal, to distinguish between Dirac and Majorana
neutrinos by exploiting the differences in the appearance spectra of the neutrino and antineutrino modes,
provided that  $g m_4 \gtrsim {\rm few}\, 0.1\, {\rm eV}$.
This might however be challenging for ND-LAr in terms of statistics, especially if $|U_{\mu 4}|^2$ is small.
A more promising avenue may be to consider a magnetized detector (such as
SAND, one of the three near detectors proposed by the DUNE collaboration)
and to reconstruct separately the electron and positron spectra in the neutrino and antineutrino modes.

Finally, while we focused in this work on the scenario of Refs.~\cite{Palomares-Ruiz:2005zbh,deGouvea:2019qre},
our results suggest that the DUNE ND-LAr detector can also efficiently probe other heavy decaying
neutrino models predicting similar signatures, such as the one of Ref.~\cite{Dentler:2019dhz}.

%%%%%%%%%%%%%%%%%%%%%%%%%%%%%%%%%%%%%%%%%

\subsection*{Acknowledgments}
We thank Guadalupe Moreno-Granados for collaboration at an early stage of this project.
The work of S.S.C. is funded by the Deutsche Forschungsgemeinschaft (DFG, German Research Foundation) Project No. 510963981.
S.S.C. also acknowledges financial support from the
LabEx P2IO (ANR-10-LABX-0038 - Project ``BSMNu'') in the framework of the ``Investissements
d'Avenir'' (ANR-11-IDEX-0003-01) managed by the Agence Nationale de la Recherche (ANR), France.
The work of S.L. is supported in part by the European Union's Horizon Europe research and innovation programme
under the Marie Sklodowska-Curie Staff Exchange grant agreement No. 101086085 -- ASYMMETRY.
S.L. also acknowledges the hospitality of the Kavli IPMU while working on this project.
O. G. M. was supported by the CONAHCyT Grant 23238 and by SNII-Mexico.
O. G. M. would like to thank Luis Delgadillo for useful discussions.

\vspace{.3cm}

This paper represents the views of the authors and should not be considered a DUNE collaboration paper.

%%%%%%%%%%%%%%%%%%%%%%%%%%%%%%%%%%%%%%%%%

%%%%%%%%%%%%%%%%%%%%%%%%%%%%%%%%%%%%%%%%%

\bibliographystyle{JHEP}
\bibliography{sterile-decay-reference}

%%%%%%%%%%%%%%%%%%%%%%%%%%%%%%%%%%%%%%%%%

%============
\end{document}